\pdfoutput=1

\documentclass[a4paper,USenglish]{lipics-v2016-modified}

\usepackage{style/allPaper}

\newcommand{\compactparagraph}[1]
{%
\vspace{0.2cm}\noindent{\textsf{\textbf{#1.}}}\quad
}%

\title{
    Liveness Verification and Synthesis:\newline New Algorithms for Recursive Programs
}

\author{Roland Meyer}
\author{Sebastian Muskalla}
\author{Elisabeth Neumann}

\affil
{
    TU Braunschweig,
    \texttt{\{roland.meyer, s.muskalla, e.neumann\}@tu-bs.de}
}

\authorrunning{R. Meyer, S. Muskalla, and E. Neumann} 

\Copyright{Roland Meyer, Sebastian Muskalla, and Elisabeth Neumann}

\newcommand {\QPA}{Q}
\newcommand {\APA}{A_{\mathit{P}}}
\newcommand {\Prio}[1]{\Omega \left(#1\right)}
\newcommand {\Prios}{\Omega \left(\QPA\right)}

\newcommand{\prefix}[1]{\text{prefix}(#1)}
\newcommand{\matchOp}{:}
\newcommand{\esol}[1]{\sigma^{e}_{#1}}
\newcommand{\towi}[2]{\tow{#1}_{#2}}

\newcommand{\derive}{\Rightarrow}

\begin{document}

\maketitle

\begin{abstract}
We consider the problems of liveness verification and liveness synthesis for recursive programs.
The liveness verification problem (LVP) is to decide whether a given $\omega$-context-free language is contained in a given $\omega$-regular language. 
The liveness synthesis problem (LSP) is to compute a strategy so that a given $\omega$-context-free game, when played along the strategy, is guaranteed to derive a word in a given $\omega$-regular language. 
The problems are known to be $\mathsf{EXPTIME}$-complete and $\mathsf{2EXPTIME}$-complete, respectively. 
Our contributions are new algorithms with optimal time complexity.
For LVP, we generalize recent lasso-finding algorithms (also known as Ramsey-based algorithms) from finite to recursive programs. 
For LSP, we generalize a recent summary-based algorithm from finite to infinite words.
Lasso finding and summaries have proven to be efficient in a number of implementations for the finite state and finite word setting.
\end{abstract}

\section{Introduction}
A major difficulty in program analysis is the combination of control and data aspects that naturally arises in programs but is not matched in the analysis: 
Control aspects are typically checked using techniques from automata theory whereas the data handling is proven correct using logical reasoning.
A promising approach to overcome this separation of techniques is a CEGAR loop recently proposed by Podelski et al.~\cite{HeizmannHoenickePodelski2010}.
The loop iteratively checks inclusions of the form $\lang{G}\subseteq \lang{B}$.
Here, $G$ is a model of the program, in the recursive setting a context-free grammar.
The automaton $B$ is a union consisting of
(1) the property of interest and (2) languages of computations that were found infeasible during the iteration by logical reasoning. 
The approach has been generalized from recursive to parallel~\cite{FKP14,LanguageRefinement} and to parameterized programs~\cite{FKP15}, from safety to liveness~\cite{FKP16}, and from verification to synthesis~\cite{HMM16}. 

We focus on the algorithmic problem behind Podelski's CEGAR loop: Inclusion checking. 
To be precise, we consider the case of recursive programs and study the problems of liveness verification and synthesis defined as follows. 
The \emph{liveness verification problem (LVP)} takes as input a context-free grammar $G$ abstracting the recursive program of interest and a B\"uchi automaton $B$ specifying the liveness property. 
The task is to check whether the $\omega$-context-free language generated by the grammar is included in the $\omega$-regular language of the automaton, $\olang{G}\subseteq \olang{B}$.
The \emph{liveness synthesis problem (LSP)} replaces the context-free grammar by a context-free game between two players: Player prover tries to establish the inclusion in an $\omega$-regular language and player refuter tries to disprove it.
The task is to synthesize a strategy $s$ such that prover is guaranteed to win all plays by following the strategy, $\olang{G@s}\subseteq \olang{B}$. 
The precise complexity of both problems is known (see below). 

Our contribution is a generalization of two recent algorithms that have proven efficient in the context of finite-state systems (for verification) and finite word languages (for synthesis) to the setting of $\omega$-languages of recursive programs. 
The study of new algorithms is motivated by the characteristics of the inclusion checks invoked in Podelski's approach: 
(1)
    The left-hand side modeling the program is substantially large. 
(2)
    The right-hand side for the specification is typically small but grows with the addition of counterexample languages. 
(3)
    These inclusion checks are invoked in an iterative fashion.
Our algorithms take into account these characteristics as follows.
First, they avoid any computation on the grammar (like intersections that would be executed in an iterative fashion).   
Second, they may terminate early if the automaton bears redundancies that may occur when languages are added. 
This early termination makes them particularly suitable for use in a refinement loop.

For the liveness verification problem LVP, we develop a \emph{lasso-finding algorithm}. 
Lasso algorithms have been proposed in~\cite{FogartyVardi} and further refined in~\cite{DBLP:conf/cav/AbdullaCCHHMV10,DBLP:conf/concur/AbdullaCCHHMV11}. 
They rely on the fact that $\omega$-regular languages can be stratified into finite unions of languages $L(\tau)L(\rho)^\omega$~\cite{Buchi1990}. 
Here, $\tau$ and $\rho$ are relations over the states of the B\"uchi automaton.
They denote the regular languages of all words that yield the prescribed state changes.
With this stratification, disproving the inclusion amounts to finding a word derivable in the $\omega$-context-free language whose representation $L(\tau')L(\rho')^\omega$ belongs to the complement of $\olang{B}$. 
Checking membership in the complement amounts to proving the absence of an accepting cycle, a lasso, in the relation $\rho'$ (when seen as a graph). 

Algorithmically, the challenge is to compute the languages $L(\tau')L(\rho')^\omega$ induced by the words derivable in the $\omega$-context-free grammar.
We view the grammar as a system of inequalities and compute the least solution over (sets of) such relations. 
The problem is to make sure that the words represented by $L(\rho')$ can be $\omega$-iterated. 
The solution is to find a lasso also in the $\omega$-context-free grammar.
To this end, we let the system of equations not only represent the non-terminal from which a terminal word is generated, but also the non-terminal from which the infinite computation will continue.
With this idea, the system is quadratic in the size of the grammar.
The height of the lattice is exponential in the number of states of the automaton.
Indeed, LVP is known to be $\mathsf{EXPTIME}$-complete~\cite{BEM97}.
The algorithm may terminate early if a language is found that disproves the inclusion.

For the liveness synthesis problem LSP, we develop a \emph{summary-based approach}.
Summaries~\cite{SharirPnueli1978,RepsHorwitzSagiv1995} represent procedures in terms of their input-output relationship on the shared memory.\footnote{Summaries resemble the aforementioned relations $\tau$ and $\rho$, see Section~\ref{Section:Liveness}.}.
Recently, summary-based analyses have been generalized to safety games by replacing relations by positive Boolean formulas of relations~\cite{HMM16}. 
We build upon this generalization and tackle the case of infinite words as follows.
In a first step, we determinize the given \Buechi automaton into a parity automaton.
The second step computes formula summaries for safety games that have the parity automaton as the right-hand side. 
Interestingly, it is sufficient to only maintain the output effect of a procedure.
In a third step, we connect the formula summaries to a parity game. 
Overall, the algorithm runs in $\mathsf{2EXPTIME}$ and indeed the problem is $\mathsf{2EXPTIME}$-complete.
The hardness is because finite games as considered in~\cite{Muscholl2005,HMM16} can be seen as a special case of LSP.
Membership in $\mathsf{2EXPTIME}$ can be shown using the techniques from \cite{Walukiewicz2001234}.

It is well known that pushdown parity games can be reduced to finite state parity games, even with a summary-like approach~\cite{Walukiewicz2001234}.
Our algorithm can therefore bee seen as a symbolic implementation of Walukiewicz's technique where formulas represent attractor information. 
Besides the compact representation, it has the advantage of being able to make use of all techniques and tools that are developed for solving fixed point equations.

\subparagraph{Related Work.}

We already mentioned the related work on the LVP and lasso finding. 
Parity games on the computation graphs of pushdown systems have been studied by Walukiewicz in~\cite{Walukiewicz2001234}. 
He reduces them to parity games on finite graphs; a technique that could also be employed to solve the LSP in $\mathsf{2EXPTIME}$.
Our algorithm which is based on solving a system of equations has the same worst-case complexity but is amenable to recent algorithmic improvements, as has been shown in~\cite{HMM16}.

Cachat~\cite{Cachat2002} considers games defined by pushdown systems in which the winning condition is reaching a configuration accepted by an alternating finite automaton once resp.\ infinitely often.
He solves them by saturating the finite automaton, in contrast to our method which uses summarization.
Although the LSP could be reduced to this type of game, the reduction has been shown to be inefficient for the case of finite games in~\cite{HMM16}.
Extensions towards higher-order systems exist~\cite{CSHORE}.

The decidability and complexity of games defined by context-free grammars has been studied by Muscholl et. al. in~\cite{Muscholl2005} and extended in \cite{Bjorklund2013,Schuster15}.
Their study considers finite games and has an emphasis on lower bounds. 
We rely on their result for the $\mathsf{2EXPTIME}$-hardness and focus on the algorithmic side.

\subparagraph{Acknowledgements.} 

We thank Prakash Saivasan and Igor Walukiewicz for discussions.

\section{$\omega$-Context-Free Languages}\label{Section:OmegaCFL}

To formulate the problem of liveness verification for recursive programs, we recall the notion of $\omega$-context-free languages~\cite{Linna1976,CohenGold1977,OFinkel2003}. 
A language $\lang \subseteq T^\omega$ of infinite words is \emph{$\omega$-context-free} if it can be written as a finite union of the form 
\[
        \lang\ =\ \bigcup_{i = 1\ldots n} V_i U_i^\omega \quad \text{with $V_i, U_i \subseteq T^*$ context-free languages of finite words.}
\]
To accept or generate $\omega$-context-free lanugages, Linna~\cite{Linna1976} as well as Cohen and Gold~\cite{CohenGold1977} define $\omega$-languages of pushdown automata and context-free grammars, respectively.
We choose the grammar-based formulation as it fits better the algebraic nature of our development.
Actually, we slightly modify the definition in~\cite{CohenGold1977} to get rid of the operational notion of repetition sets.
The correspondence will be re-established in a moment.

A \emph{context-free grammar (CFG)} is a tuple $G = (N,T,P,S)$, where $N$ is a finite set of non-terminals, $T$ is a finite set of terminals with $N \cap T = \emptyset$, $P \subseteq N \times \vartheta$ is a finite set of production rules and $S \in N$ is an initial symbol.
Here, $\vartheta = (N \cup T)^*$ denotes the set of sentential forms.
We write $X \to \eta$ if $(X, \eta) \in P$.
We assume that every non-terminal is the left-hand side of some rule.
The \emph{derivation relation $\derive$} replaces a non-terminal $X$ in $\alpha$ by the right-hand side of a corresponding rule.
Formally, $\alpha \derive \beta$ if $\alpha = \gamma X \gamma'$, $\beta = \gamma \eta \gamma'$, and there is a rule $X \to \eta\in P$.
For a non-terminal $X \in N$, we define the language $\lang{X} = \Set{ w \in T^* }{ X \derive w }$ to be the set of terminal words derivable from $X$.
The language of the grammar $\lang{G} = \lang{S}$ is the language of its initial symbol.
For sentential forms, we define $\lang{\alpha \beta} = \lang{\alpha} \lang{\beta}$, where $\lang{a} = \set{a}$ for $a \in T \cup \set{ \varepsilon }$.


Given a CFG $G$, we define its $\omega$-language $\olang{G}$ to contain all infinite words obtainable by right-infinite derivations. 
A \emph{right-infinite derivation process} $\pi$ of $G$ is an infinite sequence of rules $
        \pi
        =
        X_0 \to \alpha_0 X_1, 
        X_1 \to \alpha_1 X_2, 
        \ldots$
where the rightmost symbol of the right-hand side of each rule is the symbol on the left-hand side of the next rule, and $X_0 = S$ is the initial symbol of the grammar.
The language of such a right-infinite derivation process is the language of infinite words
\[
    \olang{\pi} = \lang{\alpha_0}  \lang{\alpha_1} \ldots
\]
Note that $\olang{\pi}$ is restricted to proper infinite words and thus does not contain $w_0\ldots w_k \varepsilon^{\omega}$.
The \emph{$\omega$-language of $G$}, denoted by $\olang{G}$, is the union over the languages $\olang{\pi}$ for all right-infinite derivation processes $\pi$ of $G$. 
The $\omega$-languages obtained in this way are precisely the $\omega$-context-free languages.
\begin{example}
    \label{ex:grammar_reqack}

    Consider the CFG $G_{ex}$ with the rules $X \rightarrow req \ Y \ ack \mid  XX$ and  $Y \ \rightarrow s \ Y \ t \mid \epsilon$ and the initial symbol $X$.
    One can show that the grammar generates the $\omega$-language $\olang{G_{ex}} = \Set{ \big(req (s^{n_i}  t^{n_i})  ack \big)^\omega }{ n_i \geq 0 \ \forall i \N}$

\end{example}

\begin{proposition}
\label{Proposition:Correspondence}
    $\lang \subseteq T^\omega$ is $\omega$-context-free if and only if $\lang = \olang(G)$ for some CFG $G$.
\end{proposition}
The same correspondence with the $\omega$-context-free languages has also been shown for the models in~\cite{Linna1976,CohenGold1977}.
Hence, the three definitions capture the same class of languages. 
This not only justifies our modification of~\cite{CohenGold1977}, it also allows us to convert a grammar into a pushdown system and vice versa in a way that is faithful wrt. $\omega$-languages.

One might ask why we do not allow intermediary infiniteness.
This would in fact decrease the expressiveness of our model.
We elaborate on this in section~\ref{Appendix:Characterization}.

The remainder of this section is dedicated to proving the proposition. 
The implication from left to right is immediate. 
For the reverse implication, we observe that the right-infinite derivation processes of a CFG can be understood as infinite paths in the $\omega$-graph, a finite graph associated with the CFG. 
From this finiteness, we can derive the structure required for an $\omega$-context-free language.
Technically, the \emph{$\omega$-graph} of $G$ is a directed graph with edges labeled by sentential forms. 
There is one vertex for each non-terminal.
Moreover, for each production rule $X \rightarrow \alpha Y$ there is an edge from $X$ to $Y$ labeled by $\alpha$.\vspace{0.1cm}
\begin{minipage}{10cm}
For the grammar $G_{ex}$ in our running example, the $\omega$-graph is depicted to the right. 
The correspondence of the derivation processes and the paths is immediate. \vspace{0.05cm}
\end{minipage}\hspace{0.4cm}
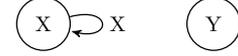
\begin{minipage}{2cm}
\scalebox{0.8}{\begin{tikzpicture}[->,>=stealth',shorten >=1pt,auto,node distance=2.8cm,
semithick]

\node[state] (A)                    {X};
\node[state] (B) [right of=A] {Y};

\path
(A) edge [loop right] node [right]  {X} (A);
\end{tikzpicture}}
\end{minipage} 

Since the $\omega$-graph is finite, every infinite path from $S$ visits some vertex $X$ infinitely often. 
We use this observation to decompose the infinite path into a finite path from $S$ to $X$ and an infinite sequence of cycles in $X$. 
This proves the next lemma.
In the statement, $\paths{Y}{Z}$ is the set of words obtained as labels of paths from $Y$ to $Z$ in the $\omega$-graph of $G$.
\begin{lemma}
\label{Lemma:DecompositionPathsAndCycles}
    Let $G$ be a CFG, then
    \[
        \textstyle
        \olang{G}\
        =\
        \bigcup_{X \in N}\ 
            \big(
                \bigcup_{ p \in \paths{S}{X}}  \lang{ p }
            \big)
            \big(
                \bigcup_{ c \in \paths{X}{X}}  \lang{ c }
                \setminus \set{ \varepsilon }
            \big)
            ^\omega
        \ .
    \]
\end{lemma}
The lemma does not yet give us the desired representation for $\olang{G}$: 
The inner unions are not finite in general, and therefore it is not clear that they define context-free languages. 
The following lemma states that this is the case.
It concludes the proof of Proposition~\ref{Proposition:Correspondence}.
\begin{lemma}
\label{Lemma:PathLanguagesContextFree}
$\bigcup_{ p \in \paths{X}{Y}}  \lang{ p}$ is context free for all non-terminals $X, Y$.
\end{lemma}

\section{Liveness Verification}
\label{Section:Liveness}

\newcommand{\varfin}[1]{\Lambda_{{#1}}}
\newcommand{\varinf}[2]{\Delta_{{#1,#2}}}
\newcommand{\varpone}[1]{\var{{#1}}}
\newcommand{\varptwo}[2]{\Psi{{#1,#2}}}
\newcommand{\solp}[2]{\sol{{#1,#2}}}
\newcommand{\solc}[1]{\sol{{#1}}}

The liveness verification problem takes as input a context-free grammar $G$ and a B\"uchi automaton $A$ and checks whether  
$\olang{G}\subseteq \olang{A}$ holds.
In the setting where $G$ is a B\"uchi automaton, recent works~\cite{FogartyVardi,DBLP:conf/cav/AbdullaCCHHMV10,DBLP:conf/concur/AbdullaCCHHMV11} have proposed so-called lasso-finding algorithms as efficient means for checking the inclusion. 
Our contribution is a generalization of lasso finding to the $\omega$-context-free case (modeling recursive rather than finite state programs).
 
A \emph{non-deterministic \Buechi automaton (NBA)} is a tuple \mbox{$A = (T, Q, q_{\mathit{init}}, Q_F, \to)$}, where $T$ is a finite alphabet, $Q$ is a finite set of states, $q_{init} \in Q$ is the initial state, $Q_F \subseteq Q$ is the set of final states, and $\to \ \subseteq Q \times T \times Q$ is the transition relation.
We write $q \overset{a}{\to} q'$ for $(q,a,q') \in\ \to$
and extend the relation to words: $q \overset{w}{\to} q'$ means there is a sequence of states starting in $q$ and ending in $q'$ labeled by $w$. 
Furthermore, we write $q \overset{w}{\to}_f q'$ if $q \tow{w} q'$ and at least one of the states in the sequence is final.
The language of infinite words $\olang{A}$ consists of all words $w\in T^\omega$ such that there is an infinite sequence of states labeled by $w$ in which infinitely many final states occur.
From now on, we use $A=(T, Q, q_{init}, Q_F, \to)$ for \Buechi automata and \mbox{$G=(N, T, P, S)$} for grammars.
Note that both use the terminal symbols $T$.

Key to our generalization are \emph{procedure summaries}. 
A procedure summary captures the changes that a procedure call may induce on the global state. 
In our setting, procedures correspond to non-terminals $X$, evaluated procedure calls to terminal words derivable from $X$, and the global state is reflected by the states of $A$. 
Hence, for every terminal word $w$ derivable from $X$ we should summarize the effect of $w$ on $A$. 
This effect are the state changes that the word may induce on the automaton.
For the set of all terminal words derivable from $X$, we thus compute the corresponding set of state changes.

We formalize procedure summaries as elements of the transition monoid~\cite{Sakarovitch2009}. 
The \emph{transition monoid} of $A$ is the monoid $\Monoid(A):=(\Boxes(A)\dotcup\set{\id}, ; , \id)$. 
The state changes on $A$ are captured by so-called \emph{boxes}, labeled relations over the states.  
The label is a flag that will be used to indicate whether the words giving rise to the relation may pass through a final state when being processed: 
\begin{align*}
\Boxes(A) :=
        \Set
        { \tboxr \in  \pwrset{  Q \times Q \times \B }   }
        { \forall q, q' \in Q: \ (q,q',1), (q,q',0) \text{ not both in } \tboxr}\ .
\end{align*}
The additional element $\id$ is the neutral element of the monoid and will play a particular role when assigning languages to boxes. 
The composition of boxes $;$ is a relational composition that remembers visits to final states.
Formally, given $\tboxr, \tboxt \in \Boxes(A)$, we define:
\begin{center}
        \begin{tabular}{lllll}
            $\tboxr;\tboxt$ & $ := $ & & $\{ (q,q',1) \ |$ &
            $ \exists q'': (q,q'',x) \in \tboxr,
            \ (q'',q',y) \in \tboxt,
            \ \max(x,y) = 1 \}$\\
            & & $\cup$ & $\{ (q,q',0)\ |$ &
            $ \exists q'': (q,q'',0) \in \tboxr,
            \ (q'',q',0) \in \tboxt,
            $\\
            & & & & $\nexists q^*:  (q,q^*,x) \in \tboxr,
            \ (q^*,q',y) \in \tboxt,
            \ \max(x,y) = 1 \}$\ .
        \end{tabular}
\end{center}
Since $\id$ is the neutral element, we have $\id;\tboxr=\tboxr;\id=\tboxr$ for all $\tboxr \in \Monoid(A)$.

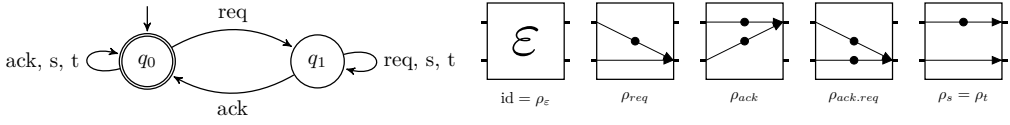
\begin{figure}[ht]
    {
        \begin{minipage}{.3\textwidth}
            \scalebox{0.8}{\begin{tikzpicture}[->,>=stealth',shorten >=1pt,auto,node distance=2.8cm,
semithick]

\node[initial above,state,accepting,above,initial text={}] (A)                    {$q_0$};
\node[state] (B) [right of=A] {$q_1$};

\path
(A) edge [bend left ] node [above]  {req} (B)
(B) edge [bend left] node [below] {ack} (A);
\path (A) edge [loop left] node [left] {ack, s, t} (A);
\path (B) edge [loop right] node [right] {req, s, t} (B);
\end{tikzpicture}}
        \end{minipage}
        \hspace*{2cm}
        \begin{minipage}{.7\textwidth}
            \scalebox{0.6}{\begin{tikzpicture}[scale=0.8]
    \pic [] (fid) [transform shape] {tbox={2}{}};
    \node [scale=4] {$\varepsilon$};
    \node [below = 1cm] {$\fid = \tbox{\varepsilon}$};
    
    \pic (req) at (3,0) [transform shape] {tbox={2}{1/2/dot, 2/2/}};
    \node [below = 1cm] at (3,0) {$\tbox{req}$};
    
    \pic (ack) at (6,0) [transform shape] {tbox={2}{2/1/dot, 1/1/dot}};
    \node [below = 1cm] at (6,0) {$\tbox{ack}$};
    
    \pic (ack) at (9,0) [transform shape] {tbox={2}{1/2/dot, 2/2/dot}};
    \node [below = 1cm] at (9,0) {$\tbox{ack.req}$};
        
    \pic (qid) at (12,0) [transform shape] {tbox={2}{1/1/dot,2/2/}};
    \node [below = 1cm] at (12,0) {$\tbox{s} = \tbox{t}$};
\end{tikzpicture}}
        \end{minipage}
    }
    \vspace*{-0.3cm}
    \caption{The automaton $A_{ex}$ and its boxes. The upper dash on each side of a box represents state $q_0$, the lower dash represents $q_1$. A dot on the dash marks that a final state has been visited.}
    \label{fig:automaton_boxes}
\end{figure}

To use boxes for checking inclusion, we have to retrieve the words represented by a box. 
A box represents the set of all words that yield the prescribed effect. 
We assign to $\tboxr\in \Boxes(A)$ the language $\lang{\tboxr}$ of all words $u\in T^+$ that satisfy $q\overset{u}{\to}q'$ for all  $(q, q', *)\in\tboxr$ and $q\overset{u}{\to}_f q'$ for all $(q, q', 1)\in \tboxr$.  
There is a $u$-labeled path from $q$ to $q'$ iff the box contains a corresponding triple (where the label is not important).
Moreover, one of the $u$-labeled paths can visit a final state iff this is required.
\[
    \lang{\tboxr}
    =
    \Set{
        u \in T^+
    }
    {
        \text{ for all } (q, q', *)\in \tboxr \colon
        q \tow{u} q'
        \text{ and for all } (q, q', 1)\in \tboxr \colon
        q \tow{u}_f q'
    }
\]
To the element $\id$ we assign the singleton language $\lang{\id}:=\set{\varepsilon}$.
The empty word cannot be lifted to an infinite word through $\omega$-iteration, and therefore has to be handled with care.
We use function $\rho: T^*\rightarrow \Monoid(A)$ to abstract a word $w\in T^*$ to the unique box $\rho_w$ representing it in the sense that $w\in \lang{\rho_w}$.
Note that $\rho_{uv}=\rho_u;\rho_v$.
This means the boxes with a non-empty language can be computed from the boxes of the letters $\rho_a$, where $\rho_a$ contains $(q, q', i)$ iff there is an $a$-labeled edge from $q$ to $q'$. We have $i=1$ iff $q$ or $q'$ is final. 
In particular, the image $\rho_{T^*}$ is precisely the set of boxes $\rho$ with $\lang{\rho}\neq \emptyset$. 
Figure~\ref{fig:automaton_boxes} illustrates the representation of words of Büchi automaton $A_{ex}$ by boxes.

\noindent%
The representation of finite words by boxes can be lifted to infinite words.
We recall two results that date back to~\cite{Vardi85}. 
The first result states that every infinite  word is contained in a composition $\lang{\tboxt}.\lang{\tboxr}^\omega$ of the languages of only two boxes. 
The proof uses Ramsey's theorem in a way similar to Theorem~\ref{Thm:LifenessVerification}, and indeed inspired our result. 
The second result states that a language $\lang{\tboxt}.\lang{\tboxr}^\omega$ is either contained in $\olang{A}$ or it is disjoint from $\olang{A}$.
It follows from the definition of box languages.  
Together, one can understand the set of languages $\lang{\tboxt}.\lang{\tboxr}^\omega$ as a finite abstraction of $T^\omega$ that is precise enough wrt. inclusion in $\lang{A}$. 
We refer to the languages $\lang{\tboxt}.\lang{\tboxr}^\omega$ included in $\lang{A}$ as the \emph{cover} of $\lang{A}$. 
\begin{lemma}
\label{lemma:sigma_omega_union_boxes}
    (1)
        For every $w \in T^\omega$ there are $\tboxt, \tboxr\in\Boxes(A)$ with $w \in \lang{\tboxt}.\lang{\tboxr}^\omega$.\\
    (2)
        Let $\tboxr, \tboxt \in \Boxes(A)$.
        We have
        $\lang{\tboxt}.\lang{\tboxr}^\omega \subseteq \olang{A}$
        or
        $\lang{\tboxt}.\lang{\tboxr}^\omega \subseteq \overline{ \olang{A} }$
        .
\end{lemma}
We compute the set of boxes summarizing the effect of the words derivable from a non-terminal as the least solution to a system of inequalities. 
The system is interpreted over the complete lattice
$\left( \pwrset{\Monoid(A)}, \subseteq \right)$.
For two sets of boxes $S, R \subseteq \Monoid(A)$, we define their composition $S;R = \Set{ \tboxt;\tboxr}{\tboxt \in S, \tboxr \in R}$ to be the set of all pairwise compositions.

There are two types of variables, $\varfin{X}$ and $\varinf{X}{Y}$ for all non-terminals $X$ and $Y$. 
The solution to a variable $\varfin{X}$ will contain the boxes for the words derivable from $X$. 
The task of $\varinf{X}{Y}$ is to additionally remember the rightmost non-terminal.  
This means we compute the boxes of all words $w$ with $X \derive^* wY$. 
The point is that the lasso-finding test has to match successive non-terminals $Y$ in the right-infinite derivation.

The inequalities for the variables $\varfin{X}$ are as follows.
For every rule $X\rightarrow \alpha$, we require
$\varfin{X}\geq \varfin{\alpha}$. 
Here, we generalize the notation $\varfin{}$ from non-terminals to sentential forms by setting $\varfin{\varepsilon}:=\set{\fid}$, 
$\varfin{a}:=\set{\tbox{a}}$, and $\varfin{\alpha\beta}:=\varfin{\alpha};\varfin{\beta}$. 
For the second set of variables, there is a base case.
For every non-terminal $Y$, we require $\varinf{Y}{Y} \geq \set{\fid}$. 
The empty word takes $Y$ to $Y$ and is represented by $\fid$. 
For every pair of non-terminals $X, Y$ and for every edge $(X, \alpha, Z)$ in the $\omega$-graph, we have the inequality
\[
    \varinf{X}{Y} \geq \varfin{\alpha};\varinf{Z}{Y} \ .
\]
To understand the requirement, assume $Z=Y$ and thus $\varinf{Y}{Y}\geq \set{\fid}$.
Then the solution to $\varinf{X}{Y}$ indeed contains the boxes of the words $w$ with $X \derive \alpha Y \derive^* wY$.
If $Z\neq Y$, we compose the solution to $\varfin{\alpha}$ with further boxes found on the way from $Z$ to $Y$.

The least solution to the above system of inequalities is computed as the least fixed point of the function on the product domain induced by the right-hand sides. 
A standard Kleene iteration~\cite{LatticesAndOrder} and more efficient methods like chaotic iteration~\cite{SeidlWilhelmHack2012} apply. 
We use $\sol{X}$ and $\sol{X, Y}$ to denote the least solution to $\varfin{X}$ and $\varinf{X}{Y}$, respectively. 
Again, we generalize the notation to sentential forms, $\sol{\alpha}$.

\begin{example}
\label{ex:eqsys_reqack}
In our running example, the system of inequalities (for $G_{ex}$ and $A_{ex}$) is
\begin{alignat*}{7}
\varfin{X}&\geq \set{\tbox{req}};\varfin{Y};\set{\tbox{ack}} 
&\hspace{0.6cm} 
\varfin{Y}&\geq \set{\tbox{s}};\varfin{Y};\set{\tbox{t}}
&\hspace{0.6cm}
\varinf{X}{X} & \geq \varfin{X};\varinf{X}{X}
&\hspace{0.6cm} 
\varinf{X}{X} & \geq\set{ \fid }\\
\varfin{X}&\geq \varfin{X};\varfin{X}&
\varfin{Y}&\geq \set{\fid}&
\varinf{X}{Y} & \geq \varfin{X};\varinf{X}{Y}&
\varinf{Y}{Y} & \geq \set{ \fid }
\end{alignat*}
The least solution is $\sol{X}=\set{\tbox{ack}}$, $\sol{Y}=\set{\fid, \tbox{s}}$, $\sol{X, X}=\set{ \fid, \tbox{ack} }$, $\sol{Y, Y}=\set{\fid}$, $\sol{X, Y}=\emptyset$, and $\sol{Y, X}=\emptyset$.

\end{example}
The following lemma states the indicated correspondence between the solution and the words in the language.
\begin{lemma}
\label{lemma:path_dfa}
$\sol{X}
        =
        \tbox{ \lang{X} }$ and  
$\solp{X}{Y}
        =
        \tbox{ \lang{ \paths{X}{Y} } }
        =
        \Set
        {
            \tbox{w}
        }
        {
            w \in \lang{p} ,
            p \in \paths{X}{Y}
        }$. 
    In particular, all occurring boxes have a non-empty equivalence class.
\end{lemma}
With the semantical results at hand, we can develop our lasso-finding algorithm.
Lassos, a notion proposed in \cite{FogartyVardi}, denote elements $\lang{\tboxt}.\lang{\tboxr}^\omega$ in the cover of $\lang{A}$ (see the discussion before Lemma~\ref{lemma:sigma_omega_union_boxes}). 
Intuitively, a pair of boxes $(\tau, \rho)$ forms a lasso if box $\rho$, when seen as a graph with the set of states as its vertex set, contains a strongly connected component that is accepting (contains a final state) and that is reachable from the first box.
\begin{definition}
A pair $(\tau$, $\rho)\in\Monoid(A)\times\Monoid(A)$ is a \emph{lasso}, if either $\rho=\id$ holds or there are states $q, q'\in Q$, a transition $(q_0, q, x) \in \tboxt$, a path from $q$ to $q'$ in $\tboxr$, and an accepting path from $q'$ to $q'$ (a loop) in $\tboxr$.
\end{definition}
\begin{minipage}{9.8cm} 
The definition is illustrated to the right. 
With the aforementioned graph-theoretic interpretation of lassos, it can be checked in linear time whether a pair of boxes actually forms a lasso.
\end{minipage}
\hspace{0.2cm}
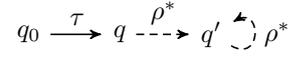
\begin{minipage}{3cm}
\begin{tikzpicture}[->,>=stealth',shorten >=1pt,auto,node distance=2.8cm,
semithick]

\node (A) {$q_0$};
\node (B) at (1.2,0) {$q$};
\node (C) [right of=B, node distance=1.2cm] {$q'$};

\path
(A) edge node [above] {$\tboxt$} (B)
(B) edge [densely dashed] node [above] {$\tboxr^*$} (C)
(C) edge [densely dashed, loop,looseness=4,in=30,out=-30] node [right] {$\tboxr^*$} (C)
;
\end{tikzpicture}
\end{minipage}

Lassos characterize the cover in the following sense.

\begin{lemma}
\label{lemma:lft}
    Let $\tboxr, \tboxt \in \Monoid(A)$ with $\lang{\tboxt}\lang{\tboxr}^\omega\neq \emptyset$. 
    Then $\lang{\tboxt}\lang{\tboxr}^\omega \subseteq \olang{A}$ holds if and only if $(\tboxt,\tboxr)$ is a lasso.   
\end{lemma}
Note that if $\tboxr=\fid$ then $\lang{\tboxt}\lang{\tboxr}^\omega=\emptyset$.
Hence, we can assume that the first case in the definition of lassos does not apply.
It is routine to check the correspondence.

Let us consider $A_{ex}$ again and choose $\tboxt = \tbox{req} = \{(q_0,q_1,1),(q_1,q_1,0)\}$ and $\tboxr = {\tbox{s}}=\{(q_0,q_0,1), (q_1,q_1,0)\}$.
The only transition in $\tboxt$ starting from initial state $q_0$ is $(q_0,q_1,1)$ and the only accepting loop in $\tboxr$ is $(q_0,q_0,1)$. However, there is no path from $q_1$ to $q_0$ in $\tboxr$. Thus, $(\tboxt, \tboxr)$ is not a lasso and $\lang{\tboxt}\lang{\tboxr}^\omega \not\subseteq \olang{A_{ex}}$.

\begin{theorem}
\label{Thm:LifenessVerification}
    The inclusion $\olang{G} \subseteq \olang{A}$ holds if and only if for every non-terminal $X \in N$, for every box $\tboxt$ in $\solp{S}{X}$ and for every box $\tboxr$ in $\solp{X}{X}$, the pair $(\tboxt, \tboxr)$ is a lasso.
\end{theorem}
One may check that using the grammar $G_{ex}$ from our running example and the automaton $A_{ex}$ from Figure~\ref{fig:automaton_boxes}, the condition is fulfilled and inclusion holds, i.e. $\olang{G_{ex}} \subseteq \olang{A_{ex}}$.

\section{Liveness Synthesis}
\label{Section:Games}

Two player games with perfect information form the theory behind synthesis problems.
In this section, we generalize a recent algorithm for solving context-free games with regular inclusion as the winning condition~\cite{HMM16} to $\omega$-context-free games with $\omega$-regular winning conditions. 
An $\omega$-context-free game is given as a context-free grammar $G=(N, T, P, S)$ where the non-terminals $N = N_\square \dotcup N_\round$ are partitioned into the non-terminals owned by player \emph{prover}~$\square$ and the ones owned by player \emph{refuter}~$\round$. 
The winning condition is defined by a \Buechi automaton $A$. 
Player $\round$ will win the game if she enforces the derivation of an infinite word not in the language of $A$. 
Player $\square$ will win the remaining plays.

Formally, the game induces a \emph{game arena}, a directed graph defined as follows. 
(1) The set of vertices is the set of all sentential forms $\vartheta = (N\dotcup T)^*$. (2) A vertex is owned by the player owning the leftmost non-terminal. 
Terminal words are owned by refuter.
(3) The edges are defined by the left-derivation relation:  
If $\alpha = w X \beta$ with $\beta \neq \varepsilon$, then $\alpha \to \gamma$ in the game arena if $\alpha \Rightarrow \gamma$ by replacing $X$. 
If $\alpha = w X$, i.e. $X$ is the leftmost and only non-terminal, then $\alpha \to \gamma$ if $\alpha \Rightarrow \gamma$ by a left-derivation using a rule $X \to \eta Y$ having a rightmost non-terminal.
A \emph{(maximal) play} of the game is a path in the game arena that is either infinite or ends in a deadlock, i.e. in a vertex that has no successor.
We think of the moves originating from vertices owned by $\square$ resp. $\round$ as chosen by prover resp. refuter.

The goal of refuter is to derive an infinite word outside $\olang{A}$, we also say that refuter plays a non-inclusion game. 
We define the infinite word derived by a play as the limit of the sequence of terminal prefixes. 
Given a sentential form $\alpha = w X \beta \in \vartheta$, 
we define its \emph{terminal prefix} to be $w \in T^*$.
An infinite play $p = \alpha_0, \alpha_1, \ldots$ of the game induces an infinite sequence of such prefixes $w_0 = \prefix{\alpha_0}, w_1 = \prefix{\alpha_1}, \ldots$, where each $w_j$ itself is a prefix of $w_{j+1}$.
Assume the words in the sequence of prefixes grow unboundedly, i.e. for any $i \in \N$, there is $j$ such that $|w_j| > i$.
The \emph{limit of the prefixes} of $p$ is the infinite word $\lim \prefix{p}$ defined by $\big(\lim \prefix{p}\big)_i = (w_j)_i$, where $w_j$ with $|w_j| > i$ is an arbitrary terminal prefix.

    An infinite play $p$ is winning with respect to the \emph{non-inclusion winning condition} if
    (1)~the prefixes of the positions in $p$ grow unboundedly, and
    (2)~$\lim \prefix{p} \not\in \olang{A}$, and
    (3)~positions of shape $wX$ occur infinitely often in $p$.
    Otherwise $p$ is winning with respect to the \emph{inclusion winning condition}.
    This is in particular the case if $p$ is finite but maximal.
Condition (1) enforces that $\lim \prefix{p}$ is a well-defined infinite word. 
Condition~(3) guarantees that it stems from a right-infinite derivation process.

Our goal is to develop an algorithm that, given a grammar and a \Buechi automaton, decides whether refuter can win non-inclusion from the initial position $S$.
Our overall strategy, following~\cite{Walukiewicz2001234}, is to reduce the problem to a finite parity game. 
The observation behind our reduction is the following.  
Each play that wins non-inclusion contains infinitely many positions of shape $wX$. 
We can therefore split the play into infinitely many parts of finite length, each starting with a position of shape $wX$. 
In a first step, we compute for every $X$ a description of all plays from $X$ to sentential forms of the shape $uY$.
In a second step, we combine the information on the finite parts into a finite parity game.

Lifting the characterization of finite plays computed in the first step to the infinite plays under study is non-trivial. 
Our approach is to determinize the given non-deterministic \Buechi automaton into a deterministic parity automaton. 
A \emph{deterministic parity automaton (DPA)} is a tuple $(\QPA,T , q_{\mathit{init}}, {\to}, \Omega)$, where $\QPA$ is a finite set of states, $q_{\mathit{init}} \in \QPA$ is the initial state, and ${\to} : \QPA \times T \to \QPA$ is the transition function.
Rather than final states, $\Omega: \QPA \rightarrow \mathbb{N}$ assigns a priority $i \in \mathbb{N}$ to each state.
We extend the transition function to words and augment it by the highest occurring priority:
We write $q\towi{w}{i}q'$ if processing $w$ starting in $q$ leads to state $q'$ and the highest priority of $q$, $q'$, and any intermediary state is $i$.
The language $\olang{\APA}$ consists of all words $w \in T^\omega$ such that the highest priority occurring infinitely often on the states in the run of $\APA$ on $w$ is even.

Non-deterministic \Buechi automata can be converted to deterministic Rabin automata~\cite{safra1988}, which in turn can be transformed to deterministic parity automata, see e.g.~\cite{piterman2007}.
\begin{theorem}[\cite{safra1988, piterman2007}]
\label{Thm:Safra}
    For an NBA $A$ with $n$ states, one can construct a DPA $\APA$ with at most $2^{\mathcal{O}(n \log n)}$ states and maximal priority $ \leq 2n+2$ so that $\olang{A}=\olang{\APA}$.
\end{theorem}
From now on, we will work with the computed DPA $\APA=(\QPA,T , q_{\mathit{init}}, {\to}, \Omega)$.


\subsection{From Context-Free Games to Formulas}
\label{Subsection:Strategies}

Our goal is to employ the characterization of inclusion games over finite words developed in~\cite{HMM16}. 
Semantically, the characterization is given as a positive Boolean formula over a finite set of atomic propositions.  
The formula captures the tree of all plays starting in a non-terminal by interpreting refuter positions as disjunctions, prover positions as conjunctions, and terminal words as atomic propositions.
Algorithmically, the formulas for all non-terminals are computed as the least solution to a system of equations. 

In the current setting,
(1)
    we have to track the priorities obtained when processing a terminal word and 
(2)
    we are given a deterministic rather than a non-deterministic automaton.
To reflect (1), we will consider as atomic propositions pairs $(p, i)$ consisting of a state $p\in \QPA$ and a priority $i\in\Prios$. 
Using (2), we define a system of equations with variables $\var{qX}$ for each state $q \in \QPA$ and each non-terminal $X \in N$. 
Intuitively, in the formula for $\var{qX}$ atomic propositions $(p,i)$ represent terminal words $w$ such that $X\derive^*w$ and $q \towi{w}{i} p$.
In the following, we define the domain and then set up the system of equations.

Let $\text{pBF}(\QPA \times \Prios)$ be the set of positive Boolean formulas over atomic propositions consisting of a state and a priority.
We will assume that the unsatisfiable formula $\ifalse$ is also contained in $\text{pBF}(\QPA \times \Prios)$. 
Conjunction $\wedge$ and disjunction $\vee$ are defined as usual.
To simplify the technical development, we evaluate operations involving $\ifalse$ on a syntactical level by using the rules $F \vee \ifalse = \ifalse \vee F = F$ and $F \wedge \ifalse = \ifalse \wedge F = \ifalse$.

Assume $F$ represents the plays from state $q$ and non-terminal $X$, and for each state $q'$ the formula $G_{q'}$ represents the plays from $q'$ and $Y$.
To obtain the formula representing the plays from $q$ and the sentential form $XY$, we can combine $F$ and the family $(G_{q'})_{q' \in \QPA}$:
A play from $XY$ to a terminal word can be decomposed into a play from $XY$ to $wY$, and a play from $wY$ to $wv$.
The first part has the same structure as a play from $X$ to $w$, and the second part is essentially a play from $Y$ to $v$ with $w$ prepended.
We think of each atomic proposition $(p,i)$ in $F$ as describing the behavior of a word $w$, i.e.\ $q \towi{w}{i} p$.
We obtain the formula imitating this behavior for $XY$ by replacing each atomic proposition $(p,i)$ in $F$ by the formula $G_p$ that describes the effect of $Y$ from $p$ on.
To reflect that the highest priority seen while processing $wv$ is the maximum of the priorities seen while processing $w$ and $v$, we will have to modify the priorities occurring in $G_p$.

We formalize the above discussion in the definitions of the composition operator $;$ on formulas and the operator~$\matchOp$ that composes one formula with a family of formulas.
Here and in the rest of the paper, we assume that $F, F', G, G' \in \text{pBF}(\QPA \times \Prios)$ are formulas, and $(G_q)_{q \in \QPA}$ and $(H_q)_{q \in \QPA}$ are families of formulas. Furthermore, $(p,i),(p',i') \in \QPA \times \Prios$ are atomic propositions and $*\in\set{\wedge, \vee}$:
\begin{align*}
    (F * F') \matchOp (G_q)_{q \in \QPA} &= F \matchOp (G_q)_{q \in \QPA} * F' \matchOp (G_q)_{q \in \QPA}
    &(p,i)\matchOp (G_q)_{q \in \QPA} &= (p,i);G_p\\
    (p,i);(G * G') &= (p,i);G * (p,i);G'
    &(p,i);(p',i') &=(p',max\set{i,i'})\ .
\end{align*}
Also here, we handle $\ifalse$ on a syntactic level by defining $F; \ifalse = \ifalse;G = \ifalse$ and $\ifalse : (G_q)_{q \in \QPA} = \ifalse$.
The case $(F * F');(p,i)$ does not occur.

We will also need to represent the terminal symbols and $\varepsilon$.
Given a state $q$ and $a \in T$, $qa$ is the formula formed by the atomic proposition $(p,i)$, where $q \towi{a}{i}p$ and $i = \max \set{ \Prio{q}, \Prio{p} }$.
To handle $\varepsilon$, we define $q \varepsilon$ to be $(q,0)$.
One might expect the second component to be $\Prio{q}$, but setting it to $0$ makes the case $\varepsilon^\omega$ (which is not an infinite word) simpler.

To guarantee that a system of equations interpreted over $\text{pBF}(\QPA \times \Prios)$ has a unique least solution, we need a partial order on the domain.
It has to have a least element and the operations have to be monotonic.
Since we deal with Boolean formulas, implication $\lleq$ is the obvious choice for the order.
Unfortunately, it is not antisymmetric, which we will tackle in a moment. 
The least element is $\ifalse$. 
Monotonicity is the following lemma.

\begin{lemma}
    \label{Lemma:Monotonicity}
    The compositions $;$ and $\matchOp$ are monotonic: If $F \lleq F'$, $G \lleq G'$, and for each $q \in Q$, $G_q \lleq G'_q$, then $F;G \lleq F';G'$ and $F \matchOp (G_q)_{q \in Q} \lleq F' \matchOp (G'_q)_{q \in Q}$.
\end{lemma}
For the solution to be computable, we have to operate on a finite domain.
Since we deal with formulas ordered by implication, we can factor them by logical equivalence.
This yields a finite domain and takes care of the missing antisymmetry. 
The order and all operations will be adapted to the domain $\text{pBF}(Q_{A_P} \times P)_{/ \lsim}$ by applying them to arbitrary representatives.
This makes $\lleq$ a partial order on the equivalence classes, and all other operations are well-defined since they were monotonic with respect to implication.

We are now ready to define the system of equations induced by $G$ and $\APA$.
To simplify the notation, we will define $\var{qa} = qa$ for $a \in T \dotcup \set{\varepsilon}$.
We extend this to sentential forms by using composition:
$\var{q\alpha\beta} = \var{q\alpha} \matchOp \left( \var{p\beta} \right)_{p \in \QPA}$.
The following lemma states that this is well-defined and not dependent on the splitting of $\alpha\beta$.

\begin{lemma}
    \label{Lemma:Assoc}
    The composition of families is associative in the following sense:
    \[
    \left(
    F
    \matchOp
    \left( G_q \right)_{q \in Q}
    \right)
    \matchOp
    \left( H_p \right)_{p \in Q}
    =
    F
    \matchOp
    \left(
    G_q
    \matchOp
    \left( H_p \right)_{p \in Q}
    \right)_{q \in Q}
    \ .
    \]
\end{lemma}
For each non-terminal $X \in$ and each state $q \in \QPA$, we have one defining equation
\[
    \var{qX} =
    \begin{cases}
        \bigwedge_{X \to \eta} \var{q\eta}\ , & X \in N_\square\ ,\\
        \bigvee_{X \to \eta} \var{q\eta}\ , & X \in N_\roundss\ .\\
    \end{cases}
\]

The resulting system of equations is solved by a standard fixed-point iteration, starting with the equivalence class of $\ifalse$ for each component.
We define $\sol{qX}$ to be the value of $\var{qX}$ in the least solution, and we extend this to sentential forms as above: $\sol{qa} = qa$ for $a \in T \dotcup \set{\varepsilon}$, $\sol{q\alpha\beta} = \sol{q\alpha} \matchOp \left( \sol{p\beta}\right)_{p \in \QPA}$. 
To show that the formula $\sol{q\alpha}$ indeed describes the behavior of all finite plays from $\alpha$, we construct strategies that are guided by the formula.

\subparagraph{Strategies.}

We fix for each equivalence class of formulas $\sol{q\alpha}$ a representative in conjunctive normal form (CNF).
(We prove that the development is independent from the choice of the representative.) 
The conjunctions correspond to the choices of prover during the play. 
The choices of refuter correspond to selecting one atomic proposition per clause. 
We formalize the selection process using the notion of choice functions. 
A \emph{choice function} on a formula $F$ is a function
\mbox{$c : F \to \QPA \times \Prios$}
selecting an atomic proposition from each clause, $c(K) \in K$ for all $K \in F$. 
We show that there is a strategy for refuter to derive at least one terminal word having one of the chosen effects on the automaton. 
In particular, the strategy will only generate finite plays. 
\begin{proposition}
\label{Proposition:FiniteStrategies}
    (1)
        Let $K$ be a clause of $\sol{q\alpha}$.
        There is a strategy $s_K$ for prover such that all maximal plays starting in $\alpha$ that conform to $s_K$ are either infinite or end in a terminal word $w$ such that $q \towi{w}{i} q'$ and $(q', i) \in K$.
    (2)
        Let $c$ be a choice function on $\sol{q\alpha}$.
        There is a strategy $s_{c}$ for refuter such that all maximal plays starting in $\alpha$ that conform to $s_{c}$ end in a terminal word $w$ with
        $q \towi{w}{i} q'$ and $(q,i) \in c(\sol{q\alpha})$.
\end{proposition}
The proof of Part (2) is a deterministic version of the analogue result in~\cite{HMM16}.
Since we do not have to guarantee termination of the play, Part (1) it is simpler.

\subparagraph{Towards Infinite Games.}

The solution of the system of equations characterizes for each non-terminal $X$ the terminal words $w$ that can be derived from $X$.
For the infinite game, we have to characterize the sentential forms $wY$ that can be derived from $X$. 
Since there may be several different non-terminals $Y$ such that a sentential form $wY$ is reachable from $X$, we store the target non-terminal in the atomic propositions. 
For $F \in \text{pBF}(\QPA \times \Prios)$ and a non-terminal $Y$, we define $F.Y$ to be the formula in $\text{pBF}(\QPA \times \Prios \times N)$ that is obtained by adding $Y$ as a third component in every atomic proposition. With $* \in \set{\vee, \wedge}$, we set
\begin{align*}
    (F * F').Y &= F.Y * F'.Y \ ,
    &(q,i).Y &= (q,i,Y) \ .
\end{align*}

For each non-terminal $X$, we collect all rules $X \to \eta Y$ with a non-terminal $Y$ as their rightmost symbol. 
We represent the behavior of $\eta$ by the previously computed formulas $\sol{q\eta}$ and attach $Y$ as described above. 
The resulting formulas are combined using disjunction or conjunction, depending on the owner of $X$. 
Given a non-terminal $X$ and a state $q \in \QPA$, we define the extended solution for $qX$ to be
\[
    \esol{qX} =
    \begin{cases}
        \bigwedge_{X \to \eta Y} \sol{q\eta}.Y, & X \in N_\square,\\
        \bigvee_{X \to \eta Y} \sol{q\eta}.Y, & X \in N_\roundss,
    \end{cases}
\]
If no rule of the shape $X \to \eta Y$ exists, the formula is $\ifalse$. 
The latter is to model that prover wins in this case, independently of who owns $X$.

\begin{proposition}
\label{Proposition:InfiniteStrategies}
    (1)
        Let $K$ be a clause of $\esol{qX}$.
        There is a strategy $s^e_K$ for prover such that all maximal plays starting in $X$ that conform to $s^e_K$ are either infinite without visiting a sentential form of shape $wY$, or they visit a sentential form of shape $wY$ such that $q \towi{w}{i} q'$ and $(q', i, Y) \in K$.
    (2)
        Let $c$ be a choice function on $\esol{qX}$.
        There is a strategy $s^e_c$ for refuter such that all maximal plays starting in $X$ that conform to $s^e_{c}$ visit a sentential form of shape $wY$ such that $q \towi{w}{i} q'$ and $(q,i,Y) \in c(\esol{qX})$.
\end{proposition}

\subsection{From Formulas to a Parity Game}
\label{Subsection:Parity}

It remains to combine the formulas for finite plays to obtain a characterization of the infinite plays.  
We model the infinite plays as an infinite sequence of alternations:
First, prover chooses a clause from the formula for $X$, which fixes her strategy for the following finite part. 
Second, refuter chooses an atomic proposition from the selected clause, which fixes the derived sentential form $wY$. 
Instead of storing the (unboundedly growing) prefixes $w$ explicitly, we only store the target non-terminal, the state transition of $\APA$ while processing $w$, and the highest priority occurring during the transition. 
Modeling the game like this leads to a parity game on a finite graph.

A \emph{parity game} $\calP = (V = V_\square \dotcup V_\roundss, E, \Omega)$ is a directed graph with an ownership partitioning of the vertices and a function $\Omega : V \to \N$ that assigns to each vertex a priority.  
We will assume that the parity game is deadlock-free. 
A maximal play is an infinite path in the graph. 
It is won by player $\square$ if the highest priority occurring infinitely often on the vertices in the play is even; won by player $\round$ otherwise. 

\begin{theorem}[Positional Determinacy of Parity Games, \cite{Zielonka1998}]
\label{Thm:Zielonka}
    Given a parity game $\calP$, there is a decomposition of the vertices
    $V = W_\square \dotcup W_\roundss$
    and there are positional strategies
    $s_\square : V_\square \to V$, $s_\roundss : V_\roundss \to V$
    such that $s_\square$ is winning from all positions in $W_\square$ and $s_\roundss$ is winning from all positions in $W_\roundss$.
\end{theorem}
%

\begin{definition}
    The parity game $\calP_{G, \APA}$ induced by the context-free grammar $G$ and the DPA $\APA$ is $(V = V_\square \dotcup V_\roundss, E, \Omega)$.
   The vertices $V_\square = \Set{ qX }{ q \in \QPA, X \in N }$ represent the formulas. 
    They are owned by prover because prover is allowed to pick a clause. 
    The vertices of refuter $V_\roundss = V^{\mathit{clause}}_\roundss \dotcup V^{\mathit{helper}}_\roundss$ are of two types. 
    Since refuter should select an atomic proposition, she owns the vertices
    $V^{\mathit{clause}}_\roundss = \Set{ qXK }{ q \in \QPA, X \in N, K \in \esol{qX}}$
    representing the clauses.
    The helper vertices
    $V^{\mathit{helper}}_\roundss = \Set{ (qXK,i,pY) }{ q \in \QPA, X \in N, K \in \esol{qX}, (p,i,Y) \in K}$
    will be used to keep track of the priority that is seen while processing the terminal prefix $w$ that is created by going from $X$ to $wY$.
    The edges connect non-terminals to clauses, and clauses to the next formula via the helper vertices:\vspace{0.1cm}
    
    \noindent
    $\begin{tabu}{l@{}l@{}l}
        E =&
        &\ \Set{\ (qX,qXK) }{q \in \QPA, X \in N, K \in \esol{qX} }\\[0.05cm]
        &\dotcup\ 
        &\ \Set{\ (qXK, (qXK,i,pY)),\ ((qXK,i,pY), pY) }{ q \in \QPA, X \in N, K \in \esol{qX}, (p,i,Y) \in K }\\[0.05cm]
        &\dotcup\ 
        &\ \Set{\ (qXK, qXK)}{ q \in \QPA, X \in N, K \in \esol{qX}, K = \emptyset }. \\[0.05cm]
    \end{tabu}$
    \noindent%
    The last part takes care of the empty clause which occurs iff the formula is equivalent to $\ifalse$.
    The priority function is zero but on the helper vertices, where it returns the priority given by the selected atomic proposition:
    $\Prio{qX} = \Prio{qXK} = 0$,
    $\Prio{(qXK,i,pY)} = i$.
\end{definition}
%
We are now able to state the correspondence between the $\omega$-context-free game of interest and the constructed parity game.

\begin{theorem}[Determinacy of $\omega$-Context-Free Games]
    Prover resp. refuter has a winning strategy for the $\omega$-regular inclusion game from $S$ iff she wins the parity game from $q_0 S$.
\end{theorem}

\begin{proof}[Proof Sketch]
    Using Theorem~\ref{Thm:Zielonka} on the positional determinacy of parity games, exactly one of the players wins the parity game from $q_0 S$, and she has a positional winning strategy.  
    We use this positional winning strategy to construct a winning strategy for the $\omega$-context-free game. 
    To this end, we establish a correspondence between the play of the parity game and the run of $\APA$ on an infinite word derived in the $\omega$-context-free grammar by following the play. 
    The key idea is that a winning strategy for the parity game for prover resp. refuter fixes clauses resp. choice functions. 
    Using these clauses resp. choice functions, we can apply Proposition~\ref{Proposition:InfiniteStrategies} to obtain a strategy for the finite part of the $\omega$-context-free game that is played until the next sentential form represented in the parity game (by a vertex) is found.
    We make this precise in Section~\ref{Appendix:Parity}.
\end{proof}

\subsection{Complexity}
\label{Subsection:Complexity}

We show that deciding whether refuter has a winning strategy for $\omega$-regular non-inclusion from position $S$ is a $\mathsf{2EXPTIME}$-complete problem.
Moreover, the algorithm presented in this section achieves this optimal time complexity.

Our proof of the lower bound works by showing that the case of finite inclusion games can be seen a special case of the problem under consideration here.
Solving finite context-free games has been shown to be a $\mathsf{2EXPTIME}$-complete problem in~\cite{Muscholl2005}.

\begin{theorem}
\label{Thm:Hardness}
    Solving $\omega$-context-free games is $\mathsf{2EXPTIME}$-hard.
\end{theorem}
We summarize the algorithm
outlined in this section:
(1)
    Construct the deterministic parity automaton $\APA$.
(2)
    Construct and solve the system of equations.
(3)
    Extend the solution $\sol{}$ to obtain $\esol{}$.
(4)
    Construct the finite parity game $\calP_G$.
(5)
    Check which player wins $\calP_G$ from $q_0S$.
    

\begin{theorem}
\label{Thm:Membership}
    Given an $\omega$-context-free game and an initial position, the algorithm outlined above decides which player wins in time
    $
        \bigO{2^{2^{|Q|^{c_1}}} \cdot 2^{|G|^{c_2}}}
    $
    for some constants $c_1, c_2 \in \N$.
\end{theorem}

\newpage

\bibliography{cited}

\newpage

\appendix 

\section{Details on Section~\ref{Section:OmegaCFL}}
\label{Appendix:Characterization}

One direction of the proof of Proposition~\ref{Proposition:Correspondence} is immediate.
We prove it in the form of the following proposition.

\begin{proposition}
\label{Proposition:OmegaContextFreeToGrammar}
    Let $\lang \subseteq T^\omega$ be a $\omega$-context-free language. Then there is a CFG $G$ such that $\lang = \olang{G}$.    
\end{proposition}
    
\begin{proof}
    We may assume
    \[\lang = \bigcup_{i = 1, ..., n} V_i U_i^\omega\]
    and there are CFGs
    \begin{align*}
        G_{V_i} &= (N_{V_i}, T, P_{V_i}, S_{V_i}),
        &G_{U_i} &= (N_{U_i}, T, P_{U_i}, S_{U_i}),
    \end{align*}
    with $V_i = \lang{G_{V_i}}$ and $U_i = \lang{G_{U_i}}$ for all $i$ such that all sets of non-terminals are pairwise disjoint.
    We construct a new grammar $G = (N, \Sigma, P, S)$ with
    $N = \set{ S } \cupdot \Set{ R_i }{ i = 1, ..., m} \cupdot \bigcupdot_{i = 1, ..., n} N_{U_i} \cupdot \bigcupdot_{i = 1, ..., n} N_{V_i}$
    and
    $P = \Set{ S \rightarrow  S_{V_i} R_i  }{ i = 1, ..., m} \cupdot \Set{ R_i \rightarrow S_{U_i}R_i }{ i = 1, ..., m} \cupdot \bigcupdot_{i = 1, ..., n} P_{U_i} \cupdot \bigcupdot_{i = 1, ..., n} P_{V_i}$.
    $\lang = \olang{G}$ is easy to see.
\end{proof}

\compactparagraph{Intermediary Infiniteness}
A grammar is supposed to generate the infinite computations of a recursive program, and a rule $X \to aYZ$ should be understood as procedure  $X$ executing action $a$, calling procedure $Y$, and after $Y$ has returned continuing with procedure $Z$.
Our restriction to the right-infinite derivations allows procedure $Z$ to run forever, but for $Y$ we only consider finite executions.
The reader may argue that we should also consider the infinite executions of $Y$.
Interestingly, our restriction to the right-infinite derivations increases the expressiveness of the language class compared to a definition that closes the $\omega$-language under intermediary infiniteness. 
The alternative definition yields a subclass of the $\omega$-context-free languages as one can always add shortened rules to a given grammar that reflect intermediary infiniteness.
In the example, one would just have to add the rule $X \to aY$ to also reflect the fact that the program may do its infinite computation without returning from procedure $Y$.
To see that the inclusion is strict, consider the language $\lang = (a^{n_i}b^{n_i})^\omega$ with $n_i \in \N$ for all~$i$.
The language containing $\lang$ would also contain $a^\omega \not\in \lang$.

\begin{proof}[Proof of Lemma~\ref{Lemma:PathLanguagesContextFree}]       
    For non-terminals $A, B \in N$ and a set $M \subseteq N$ of non-terminals, we define $\paths{A}{B}^{M}$ to be the set of all finite paths from $A$ to $B$ in the $\omega$-graph such that all occurring intermediary vertices are in $M$.
    We show that the corresponding language
    \[
        \lang{ \paths{A}{B}^M }
        = \bigcup_{ P \in \paths{A}{B}^M}  \lang{ p }
    \]
    is context-free by induction on the size of $M$. This proves that
    $\bigcup_{ p \in \paths{X}{Y}} \lang{ p }$
    is context-free since $\paths{X}{Y} = \paths{X}{Y}^N$.
    
    \noindent\textbf{\sffamily Case $M = \emptyset$:}
    All paths in $\paths{A}{B}^{\emptyset}$ have length at most one.
    If $A = B$, the corresponding language contains $\epsilon$ and all elements of the context-free languages $\lang{ \alpha }$ for all self-loops $(A, \alpha, A)$.  
    If $A \neq B$, the corresponding language contains all elements of the context-free languages $\lang{ \alpha }$ for all edges $(A, \alpha, B)$.
    Since there are only finitely many of those edges, and context-free languages are closed under finite unions, the language corresponding to $\paths{A}{B}^{\emptyset}$ is context-free.
      
    \noindent\textbf{\sffamily Case $M \neq \emptyset$:}
    Let us first consider the special case of cycles (i.e. $A = B$) and $A \in M$.
    Any cycle in which $A$ occurs as intermediary vertex can be decomposed into several cycles, such that $A$ does not occur as intermediary vertex in any of those.
    We can use this to obtain the representation
    \[
        \lang{ \paths{A}{A}^M }
        =
        \left(
            \cup_{c \in \paths{A}{A}^{M \setminus \set{A}}} L(c)
        \right) ^*
        =
        \lang{ \paths{A}{A}^{M \setminus \set{A}} }^*
    \]
    which is context-free by induction and since context-free languages are closed under Kleene-iteration.
    
    In the general case, any path $p$ from $A$ to $B$ has either no repeating intermediary vertex, i.e.\ it is simple, or there is an intermediary vertex $C$ occurring several times. In the latter case, it can be decomposed,
    \[ p = p_{AC} \ c_C \ p_{CB}\]
    where $p_{AC}$ is a path from $A$ to $C$, $p_C$ a cycle in $C$, and a $p_{CB}$ a path from $C$ to $B$.
    We can assume that $C$ does not occur as intermediary vertex in $p_{AC}$ and $p_{CB}$, and as before, we can decompose
    $c_C = c_1 c_2 \ldots c_k$
    into finitely many cycles such that $C$ does not occur in any of them as intermediary vertex.
    
    Altogether, we retrieve the representation
    \[
        \Bigg(
            \bigcup
            \limits_{
                \substack{
                    p \in \paths{A}{B}^{M} \\p \text{ simple}
               }
           }
           \lang{ p }
        \Bigg)
        \cup
        \Bigg(
            \bigcup
            \limits_{C \in N}
                \lang{ \paths{A}{C}^{M'} }
                \lang{ \paths{C}{C}^{M'} }^*
                \lang{ \paths{C}{B}^{M'} }
        \Bigg)
    \]
    where $M' = M \setminus \set{C}$.
    
    The first part is context-free since there are only finitely many simple paths. The second part is a finite union of concatenations of context-free languages (by induction and closure under Kleene-iteration).
    Altogether, this shows that
    $\lang{ \paths{A}{B}^M }$
    is context-free.
\end{proof}

\section{Details on Section~\ref{Section:Liveness}}

\begin{proof}[Proof of Lemma~\ref{lemma:path_dfa}]
    For the proof of $\sol{X} = \tbox{ \lang{X} }$, we refer to the proof of Lemma 3 in~\cite{NETYS}.
    
    It remains to prove $\solp{X}{Y} = \tbox{ \lang{ \paths{X}{Y} } }
    =
    \Set
    {
        \tbox{w}
    }
    {
        w \in \lang{p} ,
        p \in \paths{X}{Y}
    }$. 
    Assume there is a word in
    $w \in \lang{ p }, p \in \paths{X}{Y}$.
    Let $p = \alpha_0 \ldots \alpha_k$ be a decomposition of the path into its edges.
    By plugging in the inequalities into each other along the path, one can see that
    \[
        \solp{X}{Y}
        \geq
        \sol{\alpha_0};...;\sol{\alpha_k};\set{\id}
        =
        \sol{\alpha_0};...;\sol{\alpha_k}
        .
    \]
    We can write $w = w_0 ... w_m$ such that for each $i$, $w_i \in \lang{ \alpha_i }$.
    By the first part of the Lemma, we have
    $\sol{\alpha_i} = \Set{ \tbox{w} }{ w \in \lang{ \alpha_i } }$
    for each $i$, in particular $\tbox{w_i}$ in $\sol{\alpha_i}$.
    By the definition of the composition of sets of boxes and the above inequality, we then also have
    $\tbox{w} \in \solp{X}{Y}$.
    
    Assume there is a box $\tboxr$ in $\solp{X}{Y}$.
    We prove using induction that all boxes in ${\solp{X}{Y}}^{j}$ have corresponding words in $\lang{\paths{X}{Y}}$, where $\sol{}^{j}$ is the intermediary solution after the $\nth{j}$-step of Kleene iteration.
    Since $\solp{X}{Y} = {\solp{X}{Y}}^{j_0}$ for some $j_0$, this proves the claim.
    If the $\tboxr$ entered the solution in the first iteration we have $\tboxr = \fid$ and we are done.
    
    If it entered the solution in step $j > 0$, then there is some edge $(X,\alpha,Z)$ in the $\omega$-graph
    such that $\tboxr \in \sol{\alpha};{\solp{Z}{Y}}^{j-1}$,
    where $\sol{}^{j-1}$ is the solution after the $(j-1)^{th}$ iteration.
    There are boxes $\tboxt_1 \in \sol{\alpha}, \tboxt_2 \in {\solp{Z}{Y}}^{j-1}$
    such that $\tboxr = \tboxt_1;\tboxt_2$.
    By the first part of the theorem, there is a word $w_1$ such that $w_1 \in \lang{\alpha}$ with $\tbox{w_1} = \tau_1$.
    By induction, there is a word $w_2$ such that $w_2 \in \lang{\paths{Z}{Y}}$ with $\tbox{w_2} = \tboxt_2$.
    Then $w = w_1w_2$ is a word in $\lang{\paths{X}{Y}}$ with $\tbox{w} = \tbox{w_1};\tbox{w_2} = \tboxt_1;\tboxt_2 = \tboxr$.
\end{proof}

\begin{proof}[Proof of Lemma~\ref{lemma:lft}]
    Note that if $\tboxr = \fid$, then $\tboxt \tboxr^\omega = \emptyset$.
    
    Assume $(\tboxt, \tboxr)$ is a lasso, then there
    is an edge $(q_0, q,x) \in \tboxt$, a path $p$ from $q$ to some $q'$ in $\tboxr$ and a loop $c$ from $q'$ to $q'$ in $\tboxr$ such that at least one edge on the loop is labeled by one.
    Let $k$ be the length of $p$ and let $m$ be the length of $c$.
        
    Assume $w \in \lang{\tboxt} \lang{\tboxr}^\omega$, then there is a decomposition
    $w = w^{(0)} w^{(1)} \ldots$
    with $\tboxt = \tbox{w^{(0)}}$ and $\tboxr = \tbox{ w^{(1)} } = \tbox{w^{(i)}}$ for all $i > 0$.
    Then the following sequence can be refined to a run by inserting intermediary states:
    \[
        q_0
        \stackrel{ w^{(0)} } {\to}
        q
        \xrightarrow{ w^{(1)} \ldots w^{(1+k)} }
        q'
        \xrightarrow { w^{(1+k+1)} \ldots w^{(1+k+c)} }
        q'
        \xrightarrow { w^{(1+k+c+1)} \ldots w^{(1+k+2 \cdot c)} }
        \ldots
    \]
    Since at least one edge occurring in the loop is labeled by $1$, one can refine the sequence to an accepting run that visits infinitely many final states.
    This shows $w \in L^\omega(A)$.
    
    Let us now assume $w \in \lang{\tboxt}\lang{\tboxr}^\omega, w \in L^\omega(A)$,
    i.e.\ there is an accepting run of $w$ on $A$.
    Let  $w = w^{(0)} w^{(1)} ...$
    with $\tboxt = \tbox{w^{(0)}}$ and $\tboxr = \tbox{ w^{(1)} } = \tbox{w^{(i)}}$ for all $i > 0$.
    We fix an arbitrary accepting run of $A$ on $w$.
    Let $q^{(i)}$ be the state of $A$ in this run after processing $w^{(i)}$ for each $i$.
    We define $q = q^{(0)}$ and $q'$ to be the first state which occurs infinitely often in the sequence of the $q^{(i)}$.
    Let $p$ be the path from $q^{(0)}$ to $q'$ in $\tboxr$.
    There has to be an occurrence of $q'$, say $q^{(j)}$, such that there is a final state between the first occurrence of $q'$ and $q^{(j)}$.
    This proves that $\tboxr$ contains a loop $c$ from and to $q'$ in which at least one edge is labeled by $1$.
    The membership of the words in the languages of the boxes guarantees the existence of $p$ and $c$.
    The transition $(q_0,q^{(0)},*) \in \tboxt$, the path $p$ and the loop $c$ prove that $(\tboxt,\tboxr)$ is a lasso.
\end{proof}

\begin{proof}[Proof of Theorem~\ref{Thm:LifenessVerification}]
    For the implication from right to left, we show that whenever the inclusion fails, there is a non-terminal $X$, 
    a box $\tboxt\in \solp{S}{X}$, and a box $\tboxr\in \solp{X}{X}$ such that $(\tboxt, \tboxr)$ is no lasso. 
    Consider the word $w \in \olang{G}\setminus\olang{A}$.
    By definition of $\olang{G}$, there is a decomposition 
    \[ w = w^{(0)} w^{(1)} w^{(2)} ... \]
    and an infinite sequence of rules
    \[
    S \to \alpha_0 X_{1},
    X_1 \to \alpha_1 X_{2},
    \dots     
    \]
    so that $w^{(j)}\in\lang{\alpha_j}$ for all $j$.        
    Let $X$ be a non-terminal which occurs infinitely often in the sequence of the $X_i$. 
    Such an $X$ exists as there are only finitely many non-terminals.
    We create a new decomposition
    \[ w = v^{(0)} v^{(1)} v^{(2)} ... \]
    such that in the sequence of rules above, $v^{(0)}$ takes us from $S$ to $X$ for the first time, and each $v^{(j)}$ for $j > 0$ takes us from $X$ to $X$.
    
    To this decomposition, we apply Ramsey's theorem which states the following. 
    Every (undirected) infinite complete graph that has a finite edge coloring contains an infinite complete monochromatic subgraph. 
    For the application, define the labeled complete graph to have vertex set $\N$ and coloring (for all edges $\set{i,j}$ with $i < j$):
    \[
    c ( \set{ i, j})
    =
    \tbox{v_i} ; \tbox{v_{i+1}}; ... ; \tbox{v_{j-1}}
    \]
    Ramsey's theorem yields an infinite complete subgraph such that all edges have the same color. Let $S = \set{ s_0, s_1, ... }$ be the vertex set of this subgraph, with $s_0 < s_1 < \ldots$ 
    This vertex set yields a new decomposition of the word: 
    \[ w = u^{(0)} u^{(1)} u^{(2)} ... \]
    with
    \[ u^{(0)} = v^{(0)} v^{(1)} ... v^{(s_0 - 1)}\quad\text{and}\quad u^{(i)} = v^{(s_i)} v^{(s_i+1)} ... v^{(s_{i+1} - 1)}\quad \text{for all }i>0\ . \]
    Word $u^{(0)}$ takes us from $S$ to $X$ and all other $u^{(i)}$ take us from $X$ to $X$.
    We define $\tboxt = \tbox{u^{(0)}}$ and $\tboxr = \tbox{u^{(1)}}$. 
    Note that since all edges have the same color, we have $\tboxr = \tbox{u^{(i)}}$ for all $i > 0$.
    
    By construction, $\tboxt\in \solp{S}{X}$ and $\tboxr\in \solp{X}{X}$. 
    Since $w \not\in \olang{A}$, we know that $(\tboxt, \tboxr)$ is no lasso by Lemma~\ref{lemma:lft}.\\[0.2cm]
    \noindent For the implication from left to right, assume the inclusion holds, but there are $\tboxt\in\solp{S}{X}$ and $\tboxr\in\solp{X}{X}$ that do not form a lasso. 
    By Lemma \ref{lemma:path_dfa}, all boxes in $\solp{X}{Y}$ have non-empty equivalence classes. 
    We also know that $\tboxr \neq \fid$, since otherwise we would have considered the pair a lasso by definition. 
    Hence, there is an infinite word $w \in \lang{\tboxt}\lang{\tboxr}^\omega$. 
    By Lemma \ref{lemma:path_dfa} and \ref{Lemma:DecompositionPathsAndCycles}, this word is also in $\olang{G}$. 
    But by Lemma \ref{lemma:lft} and Lemma~\ref{lemma:sigma_omega_union_boxes}, $w \not\in \olang{A}$, which contradicts the assumption that the inclusion holds.
\end{proof}

\section{Details on Section~\ref{Section:Games}}

\subsection{Details on Subsection~\ref{Subsection:Strategies}}

\begin{proof}[Proof of Lemma~\ref{Lemma:Monotonicity}]
    We prove the part about $\matchOp$\ . The proof for the monotonicity of $;$ is analogous.
    The proof proceeds in phases (1) to (3) so that the claim in each phase is proven under the assumption of the claim proven in the previous phase. Let $\set{ *, \bar{*} } = \{\land,\lor \}$.
    In the following, we will use $*$ and $\bar*$ as syntactic parts of formulas as well as to connect statements in the proof.
    
    \begin{enumerate}[(1)]
        \item
        First, we prove the lemma for the case when $F,F'\in \QPA \times P$.
        In this case, $F=F'=(p,i)$ and thus $F \matchOp (G_q)_{q \in \QPA} = (p,i);G_p = F' \matchOp (G'_q)_{q \in \QPA}$.
        \item
        Next, we assume that $F'\in \QPA \times \Prios$ and $F$ is an arbitrary formula.
        We prove the statement by induction on $F$.
        
        \noindent\textbf{Base case:}    
        $F \in \QPA \times \Prios$, hence (1) proves the statement.
        
        \noindent\textbf{Induction step:}
        Let $F = F_1* F_2$. 
        Note that the Boolean formulas  $(a * b) \lleq c$ and $(a\lleq c) \mathrel{\bar{*}} (b \lleq c)$ are equivalent, called Equivalence (i) in the following.
        By the Equivalence (i), we get $(F_1\lleq F')\mathrel{\bar*}(F_2\lleq F')$. 
        Therefore, by the induction hypothesis,
        \mbox{$
            (F_1 \matchOp (G_q)_{q \in \QPA} \lleq F' \matchOp (G'_q)_{q \in \QPA})\mathrel{\bar*}(F_2 \matchOp (G_q)_{q \in \QPA}
            \lleq
            F' \matchOp (G'_q)_{q \in \QPA})
            $}.   
        This is by (i) equivalent to 
        \mbox{$
            \left( F_1 \matchOp (G_q)_{q \in \QPA} * F_2 \matchOp (G_q)_{q \in \QPA} \right)
            \lleq
            F' \matchOp (G'_q)_{q \in \QPA}$
        }. 
        By the definition of $\matchOp$, this shows
        \mbox{$
            F \matchOp (G_q)_{q \in \QPA}
            \lleq
            F' \matchOp (G'_q)_{q \in \QPA}
            $}.
        \item
        We assume that both $F, F'$ are arbitrary formulas. We prove the statement using induction on the structure of $F'$. 
        
        \noindent\textbf{Base case:}
        $F' \in \QPA \times \Prios$, hence the statement is proven by (2).
        
        \noindent\textbf{Induction step:}
        Let $F' = F_1'* F_2'$. 
        By the general equivalence of the Boolean formulas 
        $a \lleq (b* c)$ and  $(a \lleq b) * (a\lleq c)$,
        called Equivalence (ii) in the following, we get
        $(F \lleq F_1')\mathrel{*}(F \lleq F_2')$. 
        Therefore, by the induction hypothesis,
        $
        (F \matchOp (G_q)_{q \in \QPA}
        \lleq F_1' \matchOp (G'_q)_{q \in \QPA})
        \mathrel{*}
        (F\matchOp (G_q)_{q \in \QPA}
        \lleq
        F_2' \matchOp (G'_q)_{q \in \QPA})$
        holds.
        Again by (ii), we get
        $
        F \matchOp (G_q)_{q \in \QPA}
        \lleq
        (F_1' \matchOp (G'_q)_{q \in \QPA} * F_2' \matchOp (G'_q)_{q \in \QPA})
        $.  
        This is $F \matchOp (G_q)_{q \in \QPA} \lleq F' \matchOp (G'_q)_{q \in \QPA}$ by definition. 
    \end{enumerate}
    \vspace*{-0.4cm}
\end{proof}

\begin{proof}[Proof of Lemma~\ref{Lemma:Assoc}]
    The proof proceeds in phases (1) to (3) so that the claim in each phase is proven under the assumption of the claim proven in the previous phase.
    
    \begin{enumerate}[(1)]
        \item
        We first show that $((q',i) ; (q,j)) \matchOp (H_p)_{p \in \QPA} = (q',i) ; ((q,j) \matchOp (H_p)_{p \in \QPA})$.\\
        To this end, note that:
        \begin{align*}
            ((q',i) ; (q,j)) \matchOp (H_p)_{p \in \QPA} &= (q, max(i,j));H_q
            \ ,\\
            (q',i) ; ((q,j) \matchOp (H_p)_{p \in \QPA}) &= (q',i) ; ((q,j) ; H_q)
            \ .
        \end{align*}			 
        We use structural induction on $H_q$ to prove this equality.  
        
        \noindent\textbf{Base case:}
        Let $H_q=(p,k)$. Then we have 
        \begin{align*}
            &(q, max(i,j));H_q\\
            =\ & (q,max(i,j));(p,k)\\
            =\ & (p,max(i,j,k))\\
            =\ &(q',i) ; (p,max(j,k))\\
            =\ &(q',i) ; ((q,j) ; (p,k))\\
            =\ &(q',i) ; ((q,j) ; H_q)\ .
        \end{align*}
        \textbf{Induction step:}
        Let $H_q = H_1 * H_2$, with $* \in \set{\vee, \wedge}$.
        Thus,
        \begin{align*}
            &(q, max(i,j));(H_1 * H_2)\\
            =\ & (q, max(i,j));H_1 * (q, max(i,j));H_2\\
            \overset{IH}{=}\ & (q',i) ; ((q,j) ; H_1) * (q',i) ; ((q,j) ; H_2)\\
            =\ & (q',i) ; ((q,j) ; H_1 * (q,j) ; H_2)\\
            =\ & (q',i) ; (q,j) ( H_1 *  H_2)
            \ .
        \end{align*}
        \item
        We show that
        $((q',i) \matchOp (G_q)_{q \in \QPA}) \matchOp (H_p)_{p \in \QPA} = (q',i) \matchOp (G_q \matchOp (H_p)_{p \in \QPA})_{q \in \QPA}$.
        Note that
        \begin{align*}
            ((q',i) \matchOp (G_q)_{q \in \QPA}) \matchOp (H_p)_{p \in \QPA} &= ((q',i) ; G_{q'}) \matchOp (H_p)_{p \in \QPA}
            \ , \\
            (q',i) \matchOp ((G_q)_{q \in \QPA} \matchOp (H_p)_{p \in \QPA}) &= (q',i) ; (G_{q'} \matchOp (H_p)_{p \in \QPA})
            \ .
        \end{align*}
        We proceed by structural induction on $G_{q'}$.	 	
        
        \noindent\textbf{Base case:}
        $G_{q'}=(q,j)$ holds and thus (1) proves the claim.
        
        \noindent\textbf{Induction step:}
        Assume $G_q=G_1 * G_2$ for $* \in \set{\vee, \wedge}$.
        Then we can derive that  	
        \begin{align*}
            &((q',i) ; (G_1 * G_2)) \matchOp (H_p)_{p \in \QPA}\\
            =\ & ((q',i) ; G_1 * (q',i) ;G_2)) \matchOp (H_p)_{p \in \QPA} \\
            =\ & ((q',i) ; G_1) \matchOp (H_p)_{p \in \QPA} * ((q',i) ;G_2)) \matchOp (H_p)_{p \in \QPA}  \\
            \overset{IH}{=}\ & (q',i) ; (G_1 \matchOp (H_p)_{p \in \QPA}) * (q',i) ; (G_2 \matchOp (H_p)_{p \in \QPA}) \\
            =\ & (q',i) ; (G_1 \matchOp (H_p)_{p \in \QPA} *  G_2 \matchOp (H_p)_{p \in \QPA}) \\
            =\ & (q',i) ; ((G_1 * G_2) \matchOp (H_p)_{p \in \QPA})
            \ .
        \end{align*}
        \item
        Let now $(G_q)_{q \in \QPA}$ and $(H_q)_{q \in \QPA}$ be arbitrary families.
        We prove the statement by induction on the structure of $F$.
        
        \noindent\textbf{Base case:}
        $F=(q',i) \in \QPA \times \Prios$ and (2) proves the statement.
        
        \noindent\textbf{Induction step:} 
        Assume $F=F_1 * F_2$ for $* \in \set{\vee, \wedge}$. Then, we have  
        \begin{align*}
            &(F\matchOp(G_q)_{q \in \QPA}) \matchOp (H_p)_{p \in \QPA}\\
            =\ & ((F_1 * F_2)\matchOp(G_q)_{q \in \QPA}) \matchOp (H_p)_{p \in \QPA}\\
            =\ & (F_1 \matchOp(G_q)_{q \in \QPA} * F_2 \matchOp(G_q)_{q \in \QPA}) \matchOp (H_p)_{p \in \QPA}\\
            =\ & (F_1 \matchOp(G_q)_{q \in \QPA}) \matchOp (H_p)_{p \in \QPA} * (F_2 \matchOp(G_q)_{q \in \QPA})\matchOp (H_p)_{p \in \QPA} \\
            \overset{IH}{=}& F_1 \matchOp ((G_q)_{q \in \QPA} \matchOp (H_p)_{p \in \QPA}) * F_2 \matchOp ((G_q)_{q \in \QPA}\matchOp (H_p)_{p \in \QPA}) \\
            =\ & (F_1 * F_2) \matchOp ((G_q)_{q \in \QPA} \matchOp (H_p)_{p \in \QPA})\\
            =\ & F \matchOp ((G_q)_{q \in \QPA} \matchOp (H_p)_{p \in \QPA}) \text{, which proves the claim.}
        \end{align*} 
    \end{enumerate}
    \vspace*{-0.7cm}
\end{proof}  

\subparagraph{Conjunctive Normal Form.}

A formula in CNF is a conjunction of clauses, each clause being a disjunction of atomic propositions.
We use set notation and write clauses as sets of atomic propositions and formulas as sets of clauses. 
Identify $\itrue=\set{}$ and $\ifalse = \set{ \set{} }$. 

Since our formulas are negation-free, implication has a simple 
characterization. 

\begin{lemma}
\label{Lemma:Implication}
    $F \lleq G$ if and only if there is $j : G \to F$ so that $j(H) \subseteq H$ for all $H\in G$. 
\end{lemma}

\begin{proof}
    The implication from right to left is immediate.
    Assume $F\lleq G$ but there is no map $j$ as required.
    Then there is some clause $H\in G$ so that for every clause $C\in F$ we find a variable $x_C\in C$ with $x_C\notin H$.
    Consider the assignment $\nu(x_C)=\itrue$ for all $x_C$ and $\nu(y)=\ifalse$ for the remaining variables.
    Then $\nu(F)=\itrue$.
    At the same time, $\nu(G)=\ifalse$ as $\nu(H)=\ifalse$. 
    This contradicts the assumption $F\lleq G$, which means $\nu(F)=\itrue$ implies $\nu(G)=\itrue$ for every assignment $\nu$.
\end{proof}
Disjunctions and compositions can be transformed to CNF by applying distributivity.
\begin{lemma}
\label{Lemma:OperationsCNF}
    \[
        (1)\ F \vee G \lsim \Set{ K \cup H }{ K \in F, H \in G}
        ,\ \ \ 
        (2)\ F \wedge G \lsim F \cup G
        \ ,
    \]
    \[
        (3)\ F \matchOp (G_q)_{q \in Q}
        \lsim
        \bigcup_{K \in F}
        \bigcup_{\substack{z : K \to \cup_{q \in Q}G_q \\ (q,i) \mapsto H \in G_q}}
        \big\{ 
        \bigcup_{(p,j) \in K} (p,j);z((p,j))\  
        \big\}
    \]
\end{lemma}

\begin{proof}
    (1) is immediate, (2) follows from applying distributivity.
    We show (3) by structural induction on $F$.
    
    \noindent\textbf{Base case:}
    Let $F=(q,i)$. Then
    \begin{align*}
    &(q,i) \matchOp (G_q)_{q \in \QPA}\\
    =\ &(q,i);G_q\\
    =\ &\set{(q,i);K|K \in G_q}\\
    =\ &\bigcup_{\substack{z : \set{(q,i)} \to G_q \\ (q,i) \mapsto H \in G_q}} \set{(q,i);z(q,i)}
    \ .
    \end{align*}
    
    \noindent\textbf{Induction step:}\\
    We need to distinguish two cases.
    First assume $F=F_1 \land F_2$.
    Then we have
    \begin{align*}
    &(F_1 \land F_2) \matchOp (G_q)_{q \in \QPA}\\
    =\ &(F_1 \matchOp (G_q)_{q \in \QPA}) \land (F_2 \matchOp (G_q)_{q \in \QPA})\\
    \overset{(*)} {=}\ &
    \bigg( \bigcup_{K_1 \in F_1}
    \bigcup_{\substack{z_1 : K_1 \to \cup_{q \in \QPA}G_q \\ (q,i) \mapsto H \in G_q}}
    \set{ \bigcup_{(p,j) \in K_1} (p,j);z_1((p,j)) }
    \bigg)\\
    &
    \cup
    \bigg( \bigcup_{K_2 \in F_2}
    \bigcup_{\substack{z_2 : K_2 \to \cup_{q \in \QPA}G_q \\ (q,i) \mapsto H \in G_q}}
    \set{ \bigcup_{(p,j) \in K_2} (p,j);z_2((p,j)) }
    \bigg)
    \ .
    \end{align*}
    In step $(*)$, we used the induction hypothesis and part (2) of the Lemma.
    We define a function
    \[
    z: K \to \cup_{q \in \QPA}G_q,
    (q,i) \mapsto
    \begin{cases}
    z_1(q,i) \text{, if } K \in F_1 \\
    z_2(q,i) \text{, else. }
    \end{cases} 
    \]
    Using this definition, we can rewrite the last line of the equation to  
    \begin{align*}
    & \bigcup_{K \in F_1 \cup F_2}
    \bigcup_{\substack{z : K \to \cup_{q \in \QPA}G_q \\ (q,i) \mapsto H \in G_q}}
    \set{
        \bigcup_{(p,j) \in K} (p,j);z((p,j))\  
    }
    \ ,
    \end{align*}
    which proves the claim.
    
    Assume now $F=F_1 \vee F_2$. Then, 
    \begin{align*}
    &(F_1 \vee F_2) \matchOp (G_q)_{q \in \QPA}\\
    &= (F_1 \matchOp (G_q)_{q \in \QPA}) \vee (F_2 \matchOp (G_q)_{q \in \QPA})\\
    &\overset{(*)} {=} 
    \{K \cup K' | K \in S_1, K' \in S_2 \} \text{, with }\\
    &\ \ \ \ \ \
    S_1= \bigcup_{K_1 \in F_1}
    \bigcup_{\substack{z_1 : K_1 \to \cup_{q \in \QPA}G_q \\ (q,i) \mapsto H \in G_q}}
    \set{
        \bigcup_{(p,j) \in K_1} (p,j);z_1((p,j))\  
    } \\
    &\ \ \ \ \ \ 
    S_2 = 
    \bigcup_{K_2 \in F_2}
    \bigcup_{\substack{z_2 : K_2 \to \cup_{q \in \QPA}G_q \\ (q,i) \mapsto H \in G_q}}
    \set{
        \bigcup_{(p,j) \in K_2} (p,j);z_2((p,j))
    } \ .
    \end{align*}
    Therefore, 
    \begin{align*}
    &\set{K \cup K' | K \in S_1, K' \in S_2} \\
    =\ &
    \bigcup_{K_1 \in F_1} \bigcup_{K_2 \in F_2} \bigcup_{\substack{z_1 : K_1 \to \cup_{q \in \QPA}G_q \\ (q,i) \mapsto H \in G_q}}  \bigcup_{\substack{z_2 : K_2 \to \cup_{q \in \QPA}G_q \\ (q,i) \mapsto H \in G_q}}\\
    &
    \set{ \bigcup_{(p,j) \in K_1} (p,j);z_1((p,j)) }
    \cup
    \set{ \bigcup_{(p,j) \in K_2} (p,j);z_2((p,j)) }
    \end{align*}
    Using $z$ as defined above, we can rewrite this as
    \begin{align*}
    & \bigcup_{\substack{K_1 \cup K_2 \\ K_1 \in F_1, K_2 \in F_2}}
    \bigcup_{\substack{z : K_1 \cup K_2 \to \cup_{q \in \QPA}G_q \\ (q,i) \mapsto H \in G_q}}
    \set{ \bigcup_{(p,j) \in K_1 \cup K_2} (p,j);z((p,j)) }
    \ ,
    \end{align*} 
    which proves the claim.
\end{proof}
%
%
%
Towards a proof of the first part of Proposition~\ref{Proposition:FiniteStrategies}, we prove the following Lemma.

\begin{lemma}
\label{Lemma:RefineClause}
    Let $K$ be a clause of $\sol{q\alpha}$ for $\alpha = wX\beta$.\\
    (1) If $X \in N_\square$, there is $X \to \eta$ and a clause $K'$ of $\sol{qw\eta\beta}$ such that $K' \subseteq K$.\\
    (2) If $X \in N_\roundss$, for all $X \to \eta$ there is a clause $K'$ of $\sol{qw\eta\beta}$ such that $K' \subseteq K$.
\end{lemma}

\begin{proof}
    Let $q \towi{w}{i} p$, i.e. $p$ is the unique state in which $\APA$ is after processing $w$ from $q$.
    Let $F = \sol{qwX\beta}$.
    We assume that $X \to \eta_1, \ldots, X \to \eta_k$ are rules with $X$ as their left-hand side, and let $F_{\eta_i} =\sol{qw{\eta_i}\beta}$.
    \begin{enumerate}[(1)]
        \item
            By Lemma~\ref{Lemma:OperationsCNF} (3) and associativity (Lemma~\ref{Lemma:Assoc}), clause $K$ of $F$ is given by a clause $(p,i);\hat{K}$ of $\sol{qwX}$ and a function $z$ mapping this clause to $\bigcup_{q' \in \QPA} \sol{q'\beta}$.
            Since $X \in N_\square$, we have $\sol{pX} = \bigwedge_{X \to \eta_j} \sol{p\eta_j}$.
            In particular, the clause $\hat{K}$ is already a clause in $\sol{p\eta_j}$ for some $\eta_j$, and $(p,i);\hat{K}$ is a clause of $\sol{qw\eta_j}$.
            Consequently, $K$ is also a clause of $F_{q\eta_j} =\sol{qw\eta_j\beta}$.
            We may choose the rule $X \to \eta_j$ and $K' = K$.
        \item
            By Lemma~\ref{Lemma:OperationsCNF} (3) and associativity (Lemma~\ref{Lemma:Assoc}), clause $K$ of $F$ is given by a clause $(p,i);\hat{K}$ of $\sol{qwX}$ and a function $z$ mapping this clause to $\bigcup_{q' \in \QPA} \sol{q'\beta}$.
            Since $X \in N_\roundss$, we have $\sol{pX} = \bigvee_{X \to \eta_j} \sol{p\eta_j}$.
            In particular, $\hat{K} = K_1 \cup \ldots \cup K_k$, where $K_j$ is a clause of $\sol{p\eta_j}$.
            
            Let $X \to \eta_j$ be some arbitrary move.
            Note that $(p,i);K_j$ is a clause of $\sol{qw\eta_j}$, and $(p,i);K_j \subseteq (p,i);\hat{K}$.
            Consider the clause $K'$ of $F_{\eta_j} =\sol{qw{\eta_j}\beta}$ defined by $(p,i);K_j$ and the map $z$ restricted to $(p,i);K_j$.
            Note that $K = \bigcup_{(q',i') \in (p,i);\hat{K}} (q',i');z(q',i')$ and
            $
                K'
                =
                \bigcup_{(q',i') \in (p,i);K_j} (q',i');z_{\restriction_{(p,i);K_j}}(q',i')
                =
                \bigcup_{(q',i') \in (p,i);K_j}
                (q',i');z(q',i')
            $,
            so $K' \subseteq K$ holds.
    \end{enumerate}
    \vspace*{-0.4cm}
\end{proof}

\begin{proof}[Proof of Proposition~\ref{Proposition:FiniteStrategies} (1)]
    We consider the strategy $s_K$ that keeps track of a clause $K$ of the current formula.
    Initially, this clause is $K \in \sol{q\alpha}$.
    Whenever refuter makes a move $X \to \eta$, we track the clause $K'$ of the formula of the new position as in Lemma~\ref{Lemma:RefineClause} (2).
    Whenever it is our turn, we choose a rule $X \to \eta$ and the clause $K'$ of the formula of the new position as in Lemma~\ref{Lemma:RefineClause} (1).
    Note that since $K' \subseteq K$ in the Lemma, along a play $\alpha, \alpha^{(1)}, \alpha^{(2)}, \ldots$ that is conform to the strategy, we obtain a chain of clauses
    \[
        K \supseteq K^{(1)} \supseteq K^{(2)} \supseteq \ldots
        \ .
    \]
    In case the play is infinite, it has the desired property anyway.
    If it ends in a terminal word $w$, note that the formula $\sol{qw}$ is the singleton formula $\sol{qw} = \snglt{(p,i)}$, where $q \towi{w}{i} p$.
    Since we keep track of a clause of each occurring formula, the clause has to be $\set{ (p,i) }$.
    The clauses of the chain form a descending chain, so we have $\set{ (p,i) } \subseteq K$, and therefore $(p,i) \in K$.
\end{proof}
The following development aims to prove the second part of Proposition~\ref{Proposition:FiniteStrategies}.
It is mostly analogous to the development in~\cite{HMM16}, but some modifications have to be made since we consider the states of a deterministic parity automaton (plus priorities) instead of boxes for a non-deterministic automaton as atomic propositions.

The strategy $s_c$ for a choice-function $c$ is more involved, since we have to guarantee termination.
To describe how far a sentential form is away from being a terminal word, we use Kleene approximants.
Define a \emph{sequence of levels} $\lvl$ associated to a sentential form $\alpha$ to be a sequence of natural numbers of the same length.  
The formula $\sol{q\alpha}^{\lvl}$ corresponding to $\alpha$ and $\lvl$ is defined by $\sol{qa}^{i} = \snglt{qa}$ for all $a\in T\cup\set{\varepsilon}$, $\sol{qX}^{i}$ the solution to $qX$ from the $\nth{i}$ Kleene iteration,  
and $\sol{\alpha.\beta}^{\lvl.\lvl'} = \sol{q\alpha}^{\lvl} \matchOp ( \sol{q\beta}^{\lvl'} )_{q \in \QPA}$.

A choice function for $q$, $\alpha$ and $\lvl$ is a choice function on $\sol{q\alpha}^{\lvl}$.
Note that $\sol{qa}^{i}$ is independent of $i$ for terminals $qa$.
Moreover, there is an $i_0$ so that $\sol{qX}^{i_0} = \sol{qX}$ for all non-terminals $X$.
This means a choice function on $\sol{q\alpha}$ can be understood as a choice function on $\sol{q\alpha}^{i_0}$.
Here, we use a single number $i_0$ to represent a sequence $\lvl=i_0 \ldots i_0$ of the appropriate length.

By definition, $\sol{qX}^0$ is $\ifalse$ for all non-terminals, and $\ifalse$ propagates through composition by definition. 
We combine this observation with the fact that choice functions do not exist on formulas that are equivalent to $\ifalse$, because they contain an empty clause.

\begin{lemma}
\label{Lemma:ChoiceBaseCase}
    If there is a choice function for $q$, $\alpha$ and $\lvl$, then $\lvl$ does not assign zero to any non-terminal $X$ in $\alpha$.
\end{lemma}
The Lemma has an important consequence.
Consider a sentential form $\alpha$ with an associated sequence $\lvl \in 0^*$ and a choice function $c$ for $q$, $\alpha$ and $\lvl$.
Then $\alpha$ has to be a terminal word, $\alpha = w \in T^*$, $\sol{q\alpha}^{\lvl} = \snglt{(p,i)}$, where $q \towi{w}{i} p$, and the choice function has to select $(p,i)$. 
In particular, $w$ itself forms a maximal play from $w$ on, and indeed the play ends in a word whose effect is contained in the image of the choice function. 

Consider now $\alpha =wX\beta$ and $\lvl$ an associated sequence of levels.
Assume $\lvl$ assigns a positive value to all non-terminals.  
Let $j$ be the position of $X$ in $\alpha$ and let $i = \lvl_j$ be the corresponding entry of $\lvl$. 
We split $\lvl = \lvl' . \ell . \lvl''$ into the prefix for $w$, the entry $\ell$ for $X$, and the suffix for $\beta$.
For each rule $X \to \eta$, we define
$
    \lvl_\eta = \lvl' . (\ell-1) \dots (\ell-1) . \lvl''
$
to be the sequence associated  to $w \eta \beta$. 
It coincides with $\lvl$ on $w$ and $\beta$ and has entry $\ell-1$ for all symbols in $\eta$.
Note that for a terminal word, the formula is independent of the associated level, so we have
$\sol{qwX}^{\lvl'.\ell} = \sol{qwX}^{\ell}$
and
$\sol{q w \eta}^{\lvl'. (\ell-1) \ldots (\ell-1)} = \sol{q w \eta}^{\ell-1}$.

Given a choice function $c$ on a CNF-formula $F$, a choice function $c'$ on $G$ \emph{refines} $c$ if $\Set{ c'(H) }{ H \in G } \subseteq \ChosenBoxes$, denoted by $c'(G) \subseteq c(F)$.
Given equivalent formulas, a choice function on the one can be refined to a choice function on the other formula. 
Hence, we can deal with representative formulas in the following proofs.

\begin{lemma}
    \label{Lemma:ChoiceFctWellDef}
    Consider $F \lleq G$.
    For any choice function $c$ on $F$, there is a choice function $c'$ on $G$ that refines it.
\end{lemma}

\begin{proof}
    By Lemma~\ref{Lemma:Implication}, any clause $H$ of $G$ embeds a clause $j(H)$ of $F$. We can define $c'(H)$ as $c(j(H))$ to get a choice function with $c'(G)\subseteq c(F)$.
\end{proof}
We show that we can
(1)
    always refine a choice function $c$ on $\sol{q\alpha}^{\lvl}$ along the moves of prover and
(2)
    whenever it is refuter's turn, pick a specific move to refine $c$.

\begin{lemma}
\label{Lemma:RefineChoiceFct}
    Let $c$ be a choice function for $q$, $\alpha=wX\beta$ and $\lvl$.\\
    (1)
        If $X\in N_{\square}$, for all $X \to \eta$ there is a choice function $c_\eta$ for $q$, $w \eta \beta$ and $\lvl_\eta$ that refines $c$.
    (2)
        If $X\in N_{\roundss}$, there is $X \to \eta$ and a choice function $c_{\eta}$ for $q$, $w \eta \beta$ and $\lvl_\eta$ that refines $c$.
\end{lemma}

\begin{proof}
    Let $q \towi{w}{i} p$, i.e. $p$ is the unique state in which $\APA$ is after processing $w$ from $q$.
    Let $F = \sol{qwX\beta}^{\lvl}$, and for each rule $X \to \eta$, let $F_{\eta} =\sol{q\eta\beta}^{\lvl_\eta}$.
    \begin{enumerate}[(1)]
        \item
            By Lemma~\ref{Lemma:OperationsCNF} (3) and associativity (Lemma~\ref{Lemma:Assoc}), the clauses of $F$ are given by a clause $(p,i);K$ of $\sol{qwX}^{\lvl'.\ell}=\sol{qwX}^{\ell}$ and a function mapping the atomic propositions in this clause to  $\bigcup_{q' \in \QPA} \sol{q'\beta}^{\lvl''}$. 
            Similarly, the clauses of $F_{\eta}$ are given by a clause of $\sol{q w \eta}^{\ell-1}$ and a mapping from the atomic propositions to $\bigcup_{q' \in \QPA} \sol{q'\beta}^{\lvl''}$.
            We have $\sol{pX}^\ell = \bigwedge_{X \to \eta} \sol{p\eta}^{\ell-1}$.
            Since the conjunction corresponds to a union of the clause sets, Lemma~\ref{Lemma:OperationsCNF} (1), every clause of $\sol{qw\eta}^{\ell-1}$ is already a clause of $\sol{qwX}^{\ell}$. 
            Hence, the clauses of $F_\eta$ form a subset of the clauses of $F$.
            Since $c$ selects an atomic proposition from every clause of $F$, we can define the refinement $c_\eta$ on $F_\eta$ by restricting $c$.
        \item
            We show that there is a rule $X \to \eta$ and a choice function $c_\eta$ on $\sol{qw\eta \beta}^{\lvl_\eta}$ refining $c$.
            Towards a contradiction, assume this is not the case.
            Then for each rule $X \to \eta$, there is at least one clause $K_\eta''$ of $\sol{q w\eta \beta}^{\lvl_\eta}$ that does not contain an atomic proposition in the image of $c$.
            By Lemma~\ref{Lemma:OperationsCNF} (3) and associativity (Lemma~\ref{Lemma:Assoc}), $K_\eta''$ is defined by a clause $(p,i);K_\eta'$ of $\sol{w\eta}^{\ell-1}$ and a function $z_\eta$ mapping the atomic propositions from this clause to $\bigcup_{q' \in \QPA} \sol{q'\beta}^{\lvl''}$.
            
            We have $\sol{pX}^\ell = \bigvee_{X \to \eta} \sol{p\eta}^{\ell-1}$.
            A clause of $\sol{qwX}^\ell$ is thus, Lemma~\ref{Lemma:OperationsCNF} (2), of the form 
            \[
                K
                = (p,i);( \bigcup_{X \to \eta} K_\eta)
                = \bigcup_{X \to \eta} (p,i);K_\eta
                \ ,
            \]
            where each $K_\eta$ is a clause of $\sol{p\eta}^{\ell-1}$.
            We construct the clause $K' = (p,i);(\bigcup_{X \to \eta} K_\eta')$ of $\sol{qwX}^\ell$ using the $K_\eta'$ from above.
            On this clause, we define the map $z' = \bigcup_{X \to \eta} z_\eta$ that takes an atomic proposition $(p,i);(q',i) \in (p,i);K_\eta'$ and returns $z_\eta \big( (p,i);(q',i) \big)$. 
            (If an atomic proposition $(p,i);(q',i)$ is contained in $(p,i);K_\eta'$ for several $\eta$, pick an arbitrary $\eta$ among those.)
            By Lemma~\ref{Lemma:OperationsCNF} (3), $K'$ and $z'$ define a clause of $\sol{q\alpha}^{\lvl}$. 
            The choice function $c$ selects an atomic proposition $(p,i);(q',i');(q'',i'')$ out of this clause, where there is a rule $X \to \eta$ such that $(q',i') \in K_\eta'$ and
            $(q'',i'') \in z' \big( (p,i);(q',i') \big) = z_\eta \big( (p,i);(q',i') \big)$.
            This atomic proposition is also contained in $K_\eta''$, a contradiction to the assumption that no atomic proposition from $K_\eta''$ is in the image of $c$.
    \end{enumerate}
    \vspace*{-0.4cm}
\end{proof}
Notice that the sequence $\lvl_{\eta}$ is smaller than $\lvl$ in the following ordering $\natlt$ on $\N^*$. 
Given $v, w \in \N^*$, we define $v \natlt w$ if there are decompositions $v = x y z$ and $w = x i z$  so that $i > 0$ is a positive number and $y \in \N^*$ is a sequence of numbers that are all strictly smaller than~$i$.
Note that requiring $i$ to be positive will prevent the sequence $xz$ from being smaller than $x0z$, since we are not allowed to replace zeros by $\varepsilon$.

The next lemma states that $\natlt$ is well founded.
Consequently, the number of derivations $w X \beta\Rightarrow w \eta \beta$ following the strategy that refines an initial choice function will be finite.

\begin{lemma}
\label{Lemma:Wellfounded}
    $\natlt$ on $\N^*$ is  well founded with minimal elements $0^*$.
\end{lemma}

\begin{proof}
    Note that any element of $\N^*$ containing a non-zero entry is certainly not minimal, since we can obtain a smaller element by replacing any non-zero entry by $\varepsilon$. Any element of the form $0^*$ is minimal, since there is no $i$ as required by the definition of $\natlt$.
    
    Assume $v_0 \natgt v_1 \natgt \dots$ is an infinite descending chain. Let $b$ be the maximal entry of $v_0$, i.e. $b = \max_{j = 1, ..., |v_0|} v_0$, and note that no $v_l$ with $l \in \N$ can contain an entry larger than $b$ by the definition of $\natlt$.
    Therefore, we may map each $v_l$ to its Parikh image $\psi(v_l) \in \N^{b + 1}$, the vector such that $\psi(v_l)_j$ (for $j \in \zeroto{b}$) is the number of entries equal to $j$ in $v_l$.
    
    Now note that we have $\psi(v_i) > \psi(v_{i+1})$ with respect to the lexicographic ordering on $\N^{b + 1}$.
    Hence, the chain
    $\psi(v_0) > \psi(v_1) > \dots$
    is an infinite descending chain, which cannot exist since the lexicographic ordering is known to be well-founded.
\end{proof}
We have now gathered the ingredients to prove the second part of the Proposition.

\begin{proof}[Proof of Proposition~\ref{Proposition:FiniteStrategies} (2)]
    We show the following stronger claim: 
    Given any triple consisting of a sentential form $\alpha$, an associated sequence of levels $\lvl$, and a choice function $c$ for $\alpha$ and $\lvl$, there is a strategy $s_{c}$ such that all maximal plays conform to it and starting in $\alpha$ end in a terminal word $w$ with $(p,i) \in \Set{ c(K) }{ K \in \sol{q\alpha} }$, where $q \towi{w}{i} p$.
    This proves the proposition by choosing $\alpha$ and $c$ as given and $\lvl = i_0 ... i_0$, where $i_0 \in \N$ is a number such that $\sol{} = \sol{}^{i_0}$. 
    
    To show the claim, note that $\natlt$ on $\N^*$ is well founded and the minimal elements are exactly $0^*$ by Lemma~\ref{Lemma:Wellfounded}, and $\lvl_\eta \natlt \lvl$. 
    This means we can combine Lemma~\ref{Lemma:ChoiceBaseCase} and  Lemma~\ref{Lemma:RefineChoiceFct} (for the step case) into a Noetherian induction. 
    The latter lemma does not state that $\lvl_{\eta}$ assigns a positive value to each non-terminal, which was a requirement on $\lvl$. 
    This follows from Lemma~\ref{Lemma:ChoiceBaseCase} and the fact that $c_{\eta}$ is a choice function. 
    The strategy $s_{c}$ for refuter always selects the rule that affords a refinement of the initial choice function $c$.   
\end{proof}

\begin{proof}[Proof of Proposition~\ref{Proposition:InfiniteStrategies}]
    We prove both parts of the Proposition by handling the first step $X \to \eta Y$, and then using Proposition~\ref{Proposition:FiniteStrategies}.
    
    To simplify the notation in this proof, we assume that for each $\eta$ that occurs in a rule $X \to \eta Y$, the corresponding non-terminal $Y$ is unique.
    \begin{enumerate}[(1)]
        \item
            We need to distinguish two cases.
            
            \noindent\textbf{Case $X \in N_\square$:}
            Note that $\esol{qX} = \bigwedge_{X \to \eta Y} \sol{q\eta}.Y$.
            A clause of $\esol{qX}$ is of the shape $K'.Y$, where $K'$ is a clause of some $\sol{q\eta}$ with $X \to \eta Y$.
            If we play according to $s_{K'}$, the strategy constructed for $\eta$ and $K'$ in Proposition~\ref{Proposition:FiniteStrategies} (1), from $\eta Y$ on, we either end up in an infinite play that does not visit a sentential form of shape $wY$, or we end up in $wY$, with $(p,i) \in K'$, where $q \towi{w}{i} p$.
            Consequently, we have $(p,i,Y) \in K'.Y$.
            
            This means we can define $s^e_K$ by picking the rule $X \to \eta Y$, and then playing according to $s_{K'}$.
            
            \noindent\textbf{Case $X \in N_\roundss$:}
            Note that $\esol{qX} = \bigvee_{X \to \eta Y} \sol{q\eta}.Y$.
            A clause of $\esol{qX}$ is of the shape $(p,i);K'.Y$, where $K' = \bigcup_{X \to \eta Y} K_\eta$ and $K_\eta$ is a clause of $\sol{q\eta}$ for each rule $X \to \eta Y$.
            Assume prover picks a rule $X \to \eta Y$.
            Then we play according to the strategy $s_{K_\eta}$, the strategy constructed for $\eta$ and $K_\eta$ in Proposition~\ref{Proposition:FiniteStrategies} (1), from $\eta Y$ on.
            We either end up in an infinite play that does not visit a sentential form of shape $wY$, or we end up in $wvY$, with $(p_\eta,i_\eta) \in K_\eta$, where $q \towi{w}{i_\eta} p_\eta$.
            Consequently, we have $(p_\eta,i_\eta,Y) \in K_\eta.Y$.
        \item
            We again distinguish two cases, depending on who owns $X$.
            
            \noindent\textbf{Case $X \in N_\square$:}
            Note that $\esol{qX} = \bigwedge_{X \to \eta Y} \sol{q\eta}.Y$.
            A clause of $\esol{qX}$ is of the shape $K_\eta.Y$, where $K_\eta$ is a clause of some $\sol{\eta}$.
            If prover picks a rule $X \to \eta$, we can restrict the choice function $c$ to $\sol{q\eta}.Y$.
            The restricted choice function on $\sol{q\eta}.Y$ in turn induces a choice function $c_\eta$ on $\sol{q\eta}$ by ignoring the $Y$-component.
            Then we play according to $s_{c_\eta}$, the strategy constructed for $\eta$ and $c_\eta$ in Proposition~\ref{Proposition:FiniteStrategies} (2), from $\eta Y$.
            We end up in $wY$, with $(p_\eta,i_\eta) \in c_\eta( \sol{q\eta}  )$, where $q \towi{w}{i_\eta} p_\eta$.
            Consequently, we have \mbox{$(p_\eta,i_\eta,Y) \in c(\sol{qw\eta}.Y) \subseteq c(\sol{qwX}.Y)$.}
            
            \noindent\textbf{Case $X \in N_\roundss$:}
            We claim that there is a move $X \to \eta Y$ such that the given choice function induces a choice function $c_\eta$ on $\sol{q \eta}$ that is refining $c$, i.e. for each \mbox{$(p,i) \in c_\eta(\sol{q \eta})$}, there is $(p,i,Y) \in c(\esol{X})$.
            Assume this is not the case, then for each move $X \to \eta Y$ for every clause $K_\eta$ of $\sol{q \eta}$, there is no $(p_\eta,i_\eta) \in K_\eta$ such that that $(p_\eta,i_\eta,Y) \in c(\esol{qX})$.
            Consider $K = \bigcup_{X \to \eta Y} K_\eta$.
            Note that since $\esol{qX} = \bigvee_{X \to \eta Y} \sol{q\eta}.Y$,
            $K.Y$ is a clause of $\esol{qX}$.
            Therefore, the choice function $c$ selects an atomic proposition $(p_\eta,i_\eta,Y)$ from it.
            By the definition of $K$, there is a rule $X \to \eta Y$ such that $(p_\eta, i_\eta) \in K_\eta$, a contradiction to the assumption.
            
            Altogether, there is a move $X \to \eta Y$ such that $c$ induces a refinement $c_\eta$ on $\sol{q\eta}$.
            If we play according to $s_{c_\eta}$, the strategy constructed for $\eta$ and $c_\eta$ in Proposition~\ref{Proposition:FiniteStrategies} (2), from $\eta Y$, we end up in $wY$, with $(p_\eta,i_\eta) \in c_\eta( \sol{q\eta} )$, where $q \towi{w}{i_\eta} p_\eta$.
            Consequently, we have $(p_\eta,i_\eta,Y) \in c(\esol{qX})$.
    \end{enumerate}
    \vspace*{-0.4cm}
\end{proof}
\subsection{Details on Subsection~\ref{Subsection:Parity}}
\label{Appendix:Parity}

\begin{lemma}
\label{Lemma:RefineRun}
    Let $w$ be an infinite word, $w = w^{(0)} w^{(1)} \ldots$ a decomposition  into finite words,  $i^{(0)}, i^{(1)}, \ldots$ a sequence of priorities and  $q_0 = q^{(0)}, q^{(1)}, \ldots$ a sequence of states such that
    $q^{(0)} \towi{w^{(0)}}{i^{(0)}} q^{(1)} \towi{w^{(1)}}{i^{(1)}} \ldots$.
   Word $w$ is accepted by $\APA$ if and only if the highest priority  occurring infinitely often among the $i^{(j)}$ is even.
\end{lemma}

\begin{proof}
    Since $\APA$ is deterministic, there is a unique run of $\APA$ on $w$, and it can be seen as a refinement of the sequence $q^{(0)} \towi{w^{(0)}}{i^{(0)}} q^{(1)} \towi{w^{(1)}}{i^{(1)}} \ldots$.
    In particular, the highest priority $i$ occurring infinitely often in the run is the highest priority occurring infinitely often among the $i^{(j)}$.
    Assume this would not be the case. Let $j_0$ such that after $q^{(j_0)}$, no state of priority $> i$ occurs any more.
    No larger priority can occur while processing each $w^{(j)}$ for $j > j_0$, so $i$ occurs infinitely often among the $i^{(j)}$ for $j > j_0$, and no larger priority occurs at all.
    Together with the definition of the parity acceptance condition, this proves the claim.
\end{proof}

\begin{lemma}
    If $\round$ has a winning strategy for the parity game from $q_0S$, then she has a winning strategy for the grammar game from $S$.
\end{lemma}

\begin{proof}
    Let $s_\roundss$ be the positional winning strategy for the parity game from $q_0S$.
    Since $q_0S$ is owned by $\square$, each successor of $q_0 S$ has to be in the winning region, i.e.\ each vertex $q_0 S K$ corresponding to a clause $K$ of $\esol{q_0S}$.
    Since those vertices are owned by $\round$ and in her winning region, the strategy $s_\round$ selects a two-step successor $p X$ (we ignore the intermediary helper vertex) that is also in her winning region.
    Let us consider the choice function $c$ that selects the corresponding element $(p,X,i)$ in each clause $K$.
    By Proposition~\ref{Proposition:InfiniteStrategies} (2), there is a strategy for the grammar game that leads to a position $wX$ after finitely many steps, where $q_0 \towi{w}{i} p$.
    
    This position in the grammar game corresponds to the position $pX$ in the parity game, which is in refuter's winning region.
    We iterate this process to obtain both, an infinite play of the grammar game that visits positions of the shape $w'Y$ infinitely often, and a play of the parity game.
    Note that the play of the parity game is conforming to a winning strategy, and is therefore winning.
    
    We need to argue that the word generated by the infinite play of the grammar game is infinite and not in $\olang{A}$.
    The only case where the word is not infinite is that $\varepsilon$ is the only word generated infinitely often as $v$ when going from a position $w'Y$ to $w'vY'$.
    Note that by the definition of the functions $q_\varepsilon$ in the system of equations, the intermediary helper vertex $(q Y K,i,p Y')$ corresponding to the generation of $\varepsilon$ has priority $i = 0$.
    All vertices of type $q Y$ and $q Y K$ have priority $0$ as well, so $0$ would be the highest priority occurring infinitely often in the parity game, which means that the play of the parity game is not winning for $\round$.
    This is a contradiction, since the play was conform to a winning strategy for $\round$.
    
    This establishes that the word generated by the play is indeed infinite.
    Note that the word is $w_{\mathit{play}} = w^{(0)}w^{(1)} \ldots$, where each $w^{(j)}$ is the finite word generated when going from $wY$ to $ww^{(j)}Y'$.
    To argue $w_{\mathit{play}} \not\in \olang{A}$, we show that the unique run of $\APA$ on $w_{\mathit{play}}$ is not accepting.
    The run of the parity game visits infinitely many positions of type $(q X K,i,p Y)$, since the self-loops that handle empty clauses do not occur in a play won by refuter.
    Actually, the positions of this shape occurring in the play form an infinite sequence $(q^{(j)} X^{(j)} K^{(j)},i^{(j)},q^{(j+1)} X^{(j+1)})$.
    Note that $q_0 = q^{(0)} \towi{w^{(0)}}{i^{(0)}} q^{(1)} \towi{w^{(1)}}{i^{(1)}} \ldots$ forms a sequence as in Lemma \ref{Lemma:RefineRun}, and since the play was won by refuter, the largest $i^{(j)}$ occurring infinitely often has to be odd, so $w_{\mathit{play}} \not\in \olang{A}$.
\end{proof}

\begin{lemma}
    If $\square$ has a winning strategy for the parity game form $q_0S$, then she has a winning strategy for the grammar game from $S$.
\end{lemma}

\begin{proof}
%
    Let $s_\square$ be the positional winning strategy for the parity game from $q_0S$.
    Let $q_0 S K$ be the successor vertex selected by $s_\square$. Since $s_\square$ is a winning strategy, $q_0 S K$ is in the winning region.
    Note that $K$ is a clause of $\esol{q_0 S}$.
    
    By Proposition~\ref{Proposition:InfiniteStrategies} (1), there is a strategy $s_K$ for  the grammar game that either leads to an infinite play that does not visit a position of the shape $wX$, or to a position $wX$ after finitely many steps, where $(q,i,X) \in K$ and $q_0 \towi{w}{i} q$.
    In the first case, the resulting play of the grammar game is won by prover by definition.
    In the second case, note that $q_0 S K$ has a corresponding two-step successor $q X$.
    Since $q_0 S K$ was in the winning region of prover, but owned by refuter, all its successors have to be in the winning region.
    In particular, $q X$ is in the winning region of prover (and the deterministic step to the helper vertex does not matter).
    
    We iterate this process to obtain both a play of the grammar game and a play of the parity game, where the play of the parity game is conforming to a winning strategy, and is therefore winning.
    As mentioned above, the play is trivially won by prover if we do not see infinitely many positions of the shape $wX$.
    (In particular, this occurs if any of the clauses $K$ selected by $s_\square$ is empty. In this case, the corresponding play of the parity game takes the self-loop in the vertex for the clause.)
    Furthermore, plays that deadlock and plays that generate only a finite word (because $\varepsilon$ is the only word generated infinitely often) are also won by prover.
    
    Let us assume that the play of the grammar game is infinite, generates an infinite word $w_{\mathit{play}}$ and visits positions of shape $wX$ infinitely often.
    Note that $w_{\mathit{play}} = w^{(0)}w^{(1)} \ldots$, where each $w^{(j)}$ is the finite word generated when going from $wY$ to $ww^{(j)}Y'$.
    To argue $w_{\mathit{play}} \in \olang{A}$, we show that the unique run of $\APA$ on $w_{\mathit{play}}$ is accepting.
    The run of the parity game visits infinitely many positions of type $(q X K,i,p Y)$, since plays of the parity game that involve self-loops that handle empty clauses correspond to plays of the grammar gane that do not see positions of the shape $wX$ infinitely often.
    Actually, the helper vertex occurring in the play form an infinite sequence $(q^{(j)} X^{(j)} K^{(j)},i^{(j)},q^{(j+1)} X^{(j+1)})$.
    Note that $q_0 = q^{(0)} \towi{w^{(0)}}{i^{(0)}} q^{(1)} \towi{w^{(1)}}{i^{(1)}} \ldots$ forms a sequence as in Lemma \ref{Lemma:RefineRun}, and since the play was won by prover, the largest $i^{(j)}$ occurring infinitely often has to be even, so $w_{\mathit{play}} \in \olang{A}$.
\end{proof}

\subsection{Details on Subsection~\ref{Subsection:Complexity}}

\begin{proof}[Proof of Theorem~\ref{Thm:Hardness}]
    Given a context-free game $G, A$, we construct $G_\omega$ and $A_\omega$ as follows:
    We add a new symbol $\#$ to the terminal symbols.
    To obtain $G_\omega$, we add a new initial symbol $S_\omega$ and a rule $S_\omega \to S H_\omega$, where $S$ is the initial symbol of $G$, and $H_\omega$ is another fresh symbol that has a rule $H_\omega \to \# H_\omega$.
    To obtain $A_\omega$, we add a new state $q_\omega$, and for each transition that goes to a final state, we add a transition with the same input symbol to $q_\omega$.
    We add a loop $q_\omega \tow{\#} q_\omega$, and we make $q_\omega$ the only final state of $A_\omega$.
    
    Note that $\olang{G_\omega} = \lang{G} \#^\omega$, and $\olang{A} = \lang{A} \#^\omega$.
    
    Refuter wins the $\omega$-regular non-inclusion game with respect to $G_\omega, A_\omega$ from $S_\omega$ if and only if she wins the (finite) non-inclusion game $G, A$ from $S$.
    Since the size of $G_\omega$ resp. $A_\omega$ is polynomial in the size of $G$ resp. $A$, and solving (finite) non-inclusion games is $\mathsf{2EXPTIME}$-hard by~\cite{HMM16}, this proves the claim.
\end{proof}

\begin{proof}[Proof of Theorem~\ref{Thm:Membership}]
    \newcommand{\constqpa}{c_{\mathit{qpa}}}
    \newcommand{\constmaxprio}{c_{\mathit{prio}}}
    \newcommand{\constapaconstr}{c_{\mathit{det}}}
    \newcommand{\constatoms}{c_{\mathit{atom}}}
    \newcommand{\constitr}{c_{\mathit{itr}}}
    \newcommand{\constformsize}{c_{\mathit{form}}}
    \newcommand{\constfct}{c_{\mathit{fct}}}
    \newcommand{\constcomp}{c_{\mathit{comp}}}
    \newcommand{\constperitr}{c_{\mathit{peritr}}}
    \newcommand{\constconj}{c_{\wedge}}
    \newcommand{\constdisj}{c_{\vee}}
    \newcommand{\constparitystates}{c_{Q_{\mathit{parity}}}}
    \newcommand{\constparitynt}{c_{N_{\mathit{parity}}}}
    \newcommand{\conststratstates}{c_{Q_{\mathit{strat}}}}
    \newcommand{\conststratnt}{c_{N_{\mathit{strat}}}}
    \newcommand{\constextend}{c_{e}}
    
    We analyze
    (1) the time needed to construct the deterministic parity automaton,
    (2) the number of iterations needed for the fixed point-iteration,
    (3) the time consumption per iteration,
    (4) the time for extending the solution,
    (5) the time the construction of the parity game, and
    (6) the time needed to solve the parity game.
    
    Steps (2) and (3) are analogous to the proof of the corresponding result in~\cite{HMM16}.
    
    Here, $Q$ is the set of states of the \Buechi automaton $A$.
    
    \begin{enumerate}[(1)]
        \item
            As stated by Theorem~\ref{Thm:Safra}, given a NBA with $|W|$ states, one can construct a deterministic parity automaton $\APA$ with
            $
            m
            = |\QPA|
            = 2^{\bigO{|Q| \cdot \log |Q|}}
            \leq 2^{|Q|^{\constqpa}}
            $
            many states and maximal priority
            $
            2 \cdot |Q| + 2
            \leq
            |Q|^{\constmaxprio}
            $ for some $\constqpa, \constmaxprio \in \N$.
            The state set of this automaton consists of the Safra-trees for $A$, and it can be constructed in singly exponential time
            $
            2^{|Q|^{\constapaconstr}}
            $
            for some $\constapaconstr \in \N$.
        \item
            The length of any chain of strict implications of formulas over a set of $k$ atomic propositions is at most $2^k$.
            To prove this, note that modulo logical equivalence, a formula is uniquely characterized by the set of assignments such that the formula evaluates to true.
            Strict implication between two formulas implies strict inclusion between the sets.
            The statement follows since there are at most $2^k$ different truth assignments.
            We can use this to obtain that the number of iterations is bounded by $|N| \cdot m \cdot 2^k$, since the sequence of intermediary solutions is a chain in the product domain, and the height of the product domain is the height of the base domain multiplied by the number of components, i.e. $2^k$ times the number of variables $|N| \cdot m$.
            
            Each atomic proposition consists of a state in $\QPA$ and a priority in $\zeroto{2 |Q| + 2}$, i.e.
            $
            k
            = m \cdot (2 |Q| + 2)
            \leq
            2^{|Q|^{\constqpa}}
            \cdot
            |Q|^{\constmaxprio}
            \leq
            2^{|Q|^{\constatoms}}
            $.
            Altogether, the iteration needs
            \[
                |N| \cdot m \cdot 2^k
                \leq
                |N| \cdot 2^{|Q|^{\constqpa}} \cdot 2^{2^{|Q|^{\constatoms}}}
                \leq
                |N| \cdot {2^{2^{|Q|^{\constitr}}}}
            \]
            many steps for some $\constitr \in \N$.
        \item
            Let $k \leq 2^{|Q|^{\constatoms}}$ be again the number of atomic propositions.
            Every clause has at most size $k$ and there are $2^{k}$ different clauses, so every formula has size at most
            $
            k \cdot 2^{k}
            \leq
            2^{|Q|^{\constatoms}} \cdot {2^{2^{|Q|^{\constatoms}}}}
            \leq
            2^{2^{|Q|^{\constformsize}}}
            $
            for some $\constformsize \in \N$.
            
            Computing the operations used during the iteration according to Lemma~\ref{Lemma:OperationsCNF} is polynomial in the size of the formulas.
            For disjunction and conjunction, this is clear.
            For composition, we need to iterate over the at most $2^{k}$ clauses and over the at most
            \begin{align*}
                (m \cdot 2^{k} )^{k}
                &\leq
                \left(
                2^{|Q|^{\constqpa}}
                \cdot 2^{2^{|Q|^{\constatoms}}}            
                \right)^{ 2^{|Q|^{\constatoms}} }
                =
                \left(2^{|Q|^{\constqpa}}\right)^{ 2^{|Q|^{\constatoms}} }
                \cdot
                \left( 2^{2^{|Q|^{\constatoms}}} \right)^{ 2^{|Q|^{\constatoms}} }\\
                &=
                2^{{|Q|^{\constqpa}} \cdot 2^{|Q|^{\constatoms}} }
                \cdot
                2^{{2^{|Q|^{\constatoms}}} 2^{|Q|^{\constatoms}} }
                \leq
                2^{2^{|Q|^{\constfct}}}
            \end{align*}
            functions mapping atomic propositions to clauses, for some $\constfct \in \N$.
            Each clause $K$ and function $z$ determines a clause of the composition.
            To obtain its atomic proposition, we need to iterate over the at most $k$ atomic propositions $(p,i)$ of $K$ and compute $(p,i);z(p,i)$.
            To do this, we need to iterate over the at most $k$ atomic propositions $(q,j)$ of $z(p,i)$ and compute $(p,i);(q,j)$.
            Overall, to compute the relational composition of two formulas, we need
            \[
            2^{k}
            \cdot
            2^{2^{|Q|^{\constfct}}}
            \cdot
            k \cdot k \cdot (2^{|Q|^{\constqpa}} \cdot |Q|^{\constmaxprio})
            \leq
            2^{2^{|Q|^{\constcomp}}}
            \]
            steps, for some constant $\constcomp \in \N$.
            
            Per iteration, we need to carry out at most $m \cdot |G|$ conjunctions, disjunctions and relational
            compositions. Per grammar rule, we need to compute at most one conjunction or disjunction, depending on the owner of the non-terminal. For each symbol on the right-hand
            side of a grammar rule, we need to compute at most one relational composition.
            Overall, for one iteration, we need at most
            \[
            m \cdot |G| \cdot \left(
            \left(2^{2^{|Q|^{\constformsize}}}\right)^{\constconj}
            +
            \left(2^{2^{|Q|^{\constformsize}}}\right)^{\constdisj}
            +
            2^{2^{|Q|^{\constcomp}}}
            \right)
            \leq
            |G| \cdot 2^{2^{|Q|^{\constperitr}}}
            \]
            for some constants $\constconj, \constdisj, \constperitr \in \N$.
            Here, $2^{2^{|Q|^{\constcomp}}}$ is the cost of computing the relational composition earlier, and $\left(2^{2^{|Q|^{\constformsize}}}\right)^{\constconj}$ and $\left(2^{2^{|Q|^{\constformsize}}}\right)^{\constdisj}$ are rough estimations for computing conjunction and disjunction.
        \item
            If we extend the solution of the equation system, we need compute the disjunction or conjunction over at most $|G|$ many finite solutions for each non-terminal $X$ and each state $q \in \QPA$.
            Consequently, we need up to
            \[
                |N| \cdot 2^{|Q|^{\constqpa}} \cdot |G|
                \cdot
                \left(
                    \left(
                        2^{2^{|Q|^{\constformsize}}}
                    \right)^{\constconj}
                    +
                    \left(
                        2^{2^{|Q|^{\constformsize}}}
                    \right)^{\constdisj}
                \right)
                \leq
                |G| \cdot 2^{2^{|Q|^{\constextend}}}
            \]
            many steps for some constant $\constextend \in \N$.
        \item
            The parity game has $m \cdot |N|$ many vertices owned by prover.
            Furthermore, we have $m \cdot |N| \cdot 2^{k'}$ many vertices for the clauses owned by refuter, where $k' = m \cdot (2 |Q| + 2) \cdot |N|$ is the number of atomic propositions in the extended equation system.
            Additionally, there are up to $m \cdot |N| \cdot 2^{k'} \cdot k'$ many helper vertices also owned by refuter.
            The number of edges is polynomial in the number of vertices, and the highest occurring priority is again $2 |Q| + 2$.
            Altogether, the size of the parity game is at most
            \[
                m \cdot |N| + m \cdot |N| \cdot 2^{k'} + m \cdot |N| \cdot 2^{k'} \cdot k'
                \leq
                2^{2^{|Q|^{\constparitystates}}} \cdot 2^{|N|^{\constparitynt}}
            \]          
            for some constants $\constparitystates, \constparitynt \in \N$.
            Furthermore, the parity game can be constructed in a time that is polynomial in its size.
            If we choose the constants to be sufficiently high, the constructing the parity game is possible within the same bound.
        \item
            In general, solving parity games is in $\mathsf{NP} \cap \mathsf{coNP}$, so we would not expect to have a deterministic $\mathsf{2EXPTIME}$-algorithm that solves our parity game of doubly-exponential size.
            
            Fortunately, the number of nodes by prover is only singly exponential, namely \mbox{$m \cdot |N| \leq 2^{|Q|^{\constqpa}} \cdot |N|$.}
            By Theorem~\ref{Thm:Zielonka}, prover wins the parity game if and only if she has a positional winning strategy $s_\square$.
            A positional strategy is a function that takes a vertex $pX$ and returns one of the $2^{k'}$-many clauses of $\esol{pX}$.
            Therefore, there are only up to
            \[
                (2^{k'})^{2^{|Q|^{\constqpa}} \cdot |N|}
                \leq
                2^{2^{|Q|^{\constqpa}} \cdot (2 |Q| + 2) \cdot |N| \cdot 2^{|Q|^{\constqpa}} \cdot |N|}
                \leq
                2^{2^{|Q|^{\conststratstates}}} \cdot 2^{|N|^{\conststratnt}}
            \]
            many such strategies, for some constants $\conststratstates, \conststratnt \in \N$.
            
            After some strategy is fixed, checking whether it is winning is polynomial in the size of the game.
            We drop all edges originating in vertices owned by prover that were not selected by the winning strategy.
            In the resulting graph, we need to check whether refuter can win.
            In this case, the strategy would not be winning.
            To do this, we have to check whether there is a vertex $v$ reachable from the initial vertex $q_0S$ such that there is a cycle from/to $v$ such that the highest priority occurring on an edges in the cycle is odd.
            
            If we choose the constants $\conststratstates, \conststratnt$ to be sufficiently large, one can iterate over all strategies and check whether they are winning in $2^{2^{|Q|^{\conststratstates}}} \cdot 2^{|N|^{\conststratnt}}$ many steps.
    \end{enumerate}
    \mbox{}
    Altogether we need
    \begin{align*}
        2^{|Q|^{\constapaconstr}}
        &+
        \left(
        |N| \cdot {2^{2^{|Q|^{\constitr}}}}
        \right)
        \cdot
        \left(
        |G| \cdot 2^{2^{|Q|^{\constperitr}}}
        \right)
        +
        |G| \cdot 2^{2^{|Q|^{\constextend}}}\\
        &+
        2^{2^{|Q|^{\constparitystates}}} \cdot 2^{|N|^{\constparitynt}}
        +
        2^{2^{|Q|^{\conststratstates}}} \cdot 2^{|N|^{\conststratnt}}
        \leq
        2^{2^{|Q|^{c_1}}} \cdot 2^{|G|^{c_2}}
    \end{align*}
    many steps for some constants $c_1, c_2 \in \N$.
    Here, we used the rough estimation $|N| \leq |G|$.
\end{proof}

\end{document}